\documentclass{aa}  

\usepackage{verbatim}
\usepackage{graphicx}
\usepackage{multirow}
\usepackage{txfonts}
\usepackage[colorlinks=true,citecolor=blue]{hyperref}
\newcommand{\comm}[1]{\textcolor{black}{#1}}

\begin{document} 
   \title{Hamilton's Object revisited: A challenging source redshift of a strong lensing configuration}
\titlerunning{Hamilton's Object Revisited}

   \author{Jenny Wagner\inst{1,2,3} and
          Emilio E.~Falco\inst{4}}

   \institute{Helsinki Institute of Physics, P.O. Box 64, FI-00014 University of Helsinki, Finland
        \and 
        Academia Sinica Institute of Astronomy and Astrophysics, 11F of AS/NTU Astronomy-Mathematics Building, No.1, Sec.~4, Roosevelt Rd, Taipei 106216, Taiwan, R.O.C.
        \and 
        Bahamas Advanced Study Institute and Conferences, 4A Ocean Heights, Hill View Circle, Stella Maris, Long Island, The Bahamas\\
         \email{wagner@asiaa.sinica.edu.tw}
        \and Center for Astrophysics, Harvard \& Smithsonian (Ret.), 60 Garden St., Cambridge, MA 02138, USA \\
        \email{eefalco@gmail.com}
}

   \date{Received XXX; accepted YYY}

  \abstract
{Low-resolution spectrographs used to have difficulties in determining redshifts of galaxies at $z\approx1$ and $z\approx3$. Spectral emission and absorption lines of magnesium and iron redshifted to $z\approx1$ fall close to hydrogen, silicon, and oxygen lines at $z\approx3$. Here, we demonstrate that even with modern, integrated field unit spectrographs, this task remains challenging. Hamilton's Object, a blue star-forming galaxy, gravitationally lensed into three multiple images by the galaxy cluster SDSS J223010.47-081017.8, is such a case. Using the Blue Keck Cosmic Web Imager (KCWI) its redshift was determined as $z=0.82$, while its MOIRCS spectrum hinted at $z=3.201$. To resolve the ambiguity, we completely reanalyse the Blue KCWI spectra of all three multiple images including the star-forming region in the outskirts. We employ a new data reduction pipeline, \texttt{PypeIt}, signal enhancement, and line fitting by \texttt{Python}-routines. The reevaluation confirms the previous result based on six absorption features, $z=0.820 \pm 0.001$, and four emission features, $z=0.821 \pm 0.002$. The alternative $z=3.199\pm 0.003$, based on six absorption and two emission lines, is a worse fit, also compared to other spectra. Moreover, we find the MOIRCS spectrum inconclusive: observations cover two of three multiple images, with the slit for image~C only covering its central bulge; furthermore the pixel-to-wavelength calibration requires a nightsky emission-line calibration due to a missing calibration arc lamp. 
New MOIRCS observations are needed to verify that Hamilton's Object has the smallest separation \comm{in angular diameter distance} between lensing cluster and source galaxy among \comm{the} known cluster-scale strong lenses. }

\maketitle

\nolinenumbers  
\section{Introduction}
\label{sec:introduction}

In recent years, strong gravitational lensing by galaxy clusters has become increasingly valuable to constrain the mass distribution in galaxy clusters and to study the properties of the background source galaxies at high redshift. 
The reason is that many multiple-image configurations have been discovered that show a high level of detail in their surface brightness profiles; see, for instance, the selected examples listed in Table~\ref{tab:related_clusters} and \cite{bib:Bayliss2011}.

\begin{table*}[ht]
 \caption[]{\label{tab:related_clusters}Multiple-image configurations that show a great amount of detail in their surface brightness profiles comparable to the one studied in this work.}
\begin{tabular}{lcccccl}
 \hline \hline
  Cluster name & sys. & $n_\mathrm{im}$ & $z_\mathrm{d}$ & $z_\mathrm{s}$ & Ref. & Remarks\\
 \hline
A370 & 2 &5 & 0.375 & 0.725 & 1 & giant arc, one of the first ever discovered\\
A3827 & 1 & 5 & 0.099 & 1.24 & 2 & almost Einstein ring\\ 
CL0024 & 1 & 5 & 0.39& 1.675 & 3 & high degree of axisymmetry, merger along the line of sight\\
DESI-090.9854 & 3 & 3 & 0.49 & 1.166 & 4 & ``Carousel Lens'', naked cusp configuration of images\\
DESI-090.9854 & 4 & 4 & 0.49 & 1.432 & 4 & Einstein cross\\
GAL-CLUS-022058 & 1 & 4 & 0.36 & 1.4796 & 5 & ``Molten Ring", almost Einstein ring\\
MACSJ0416 &12 & 3 & 0.396 & 0.9392 & 6 & similar fold configuration like the one in this work \\
MACSJ1149 & 1 & 3 & 0.542 & 1.489 & 7 & host of SN Refsdal\\
RCS2 032727 & 1 & 5 & 0.564 & 1.7 & 8 & giant arc\\
RXJ0437 & 1 & 4 & 0.285 & 2.9732 & 9 & exotic lens\\
RXJ0437 & 2 & 4 & 0.285 & 1.9722 & 9 & containing three hyperbolic umbilic lensing configurations\\
SDSSJ1226+2152 & 10 & 3 & 0.4358 & 2.9233 & 10 & similar fold like the one in this work\\
SDSS J1226+2149 & 30 & 3 & 0.4358 & 1.6045 & 10 & similar fold like the one in this work \\
\hline
SDSS J223010  & 1 & 3 & 0.526 & 0.82 & 11 & analysed in this work \\
SDSS J223010  & 1 & 3 & 0.6214 & 3.201 & 12 & analysed in this work \\
\hline
\end{tabular}
\tablefoot{The first column shows the cluster name, the second the multiple-image system as defined in the paper referenced in the sixth column. The third column contains the number of multiple images belonging to this background source discovered so far, then $z_\mathrm{d}$ and $z_\mathrm{s}$ denote the redshifts of the cluster and the source, respectively.}
\tablebib{(1)~\citet{bib:Lagattuta2019};
(2) \citet{bib:Lin2023}; (3) \citet{bib:Wagner_cluster0024}; (4) \citet{bib:Sheu2024};
(5) \citet{bib:Diaz2021}; (6) \citet{bib:Vanzella2021}, \citet{bib:Caminha2017}; (7) \citet{bib:Grillo2016};
(8) \citet{bib:Sharon2012}; (9) \citet{bib:Lagattuta2023}; (10) \citet{bib:Bayliss2011}, \cite{bib:Sharon2022}; 
(11) \citet{bib:Griffiths2021}; (12) \citet{bib:Ebeling2025}; $z_\mathrm{d}$ is assumed to be the peak in the redshift distribution of the biggest merging clump in the cluster.
}
\end{table*}

As has been shown in a series of works \cite{bib:Wagner1, bib:Wagner2, bib:Wagner3, bib:Wagner4, bib:Wagner5, bib:Wagner6, bib:Wagner7}, local properties of the light-deflecting mass can be directly extracted from highly detailed multiple images. 
They do not need to fit any global mass density profile to describe the lens, as lens-modelling tools such as \texttt{lenstool} (\citet{bib:Jullo2007} and \citet{bib:Jullo2009}) or \texttt{Grale} (\citet{bib:Liesenborgs2006}, \citet{bib:Liesenborgs2010}, and \citet{bib:Liesenborgs2020}) do. 
The local lens properties are the maximum information that all lens models agree upon. Model assumptions, such as light-traces mass, take over in the reconstruction of the light-deflecting mass density with increasing distance from the multiple-image positions, as has been shown by \cite{bib:Wagner_cluster0024} and \cite{bib:Lin2023} for cluster-scale lenses and in \cite{bib:Wagner_quasar} for galaxy-scale lenses. 
Hence, such detailed multiple-image configurations provide the grounds to compare lens model-based reconstructions of cluster mass densities with each other and investigate their predictive power in regions devoid of multiple images. 

One prominent example for such an analysis was done for the galaxy cluster Abell~3827. 
First, lens model-based mass maps created in \cite{bib:Williams2011} showed an offset between the surface brightness profiles of the central galaxies and their total mass density as reconstructed from the multiple image configuration. 
With increasingly detailed observations, obtained by \cite{bib:Massey2015} and \cite{bib:Massey2018}, this offset could be shown to be negligible, which was confirmed in \cite{bib:Lin2023} without employing any lens modelling. 

The triple-image configuration SDSS J223010.47-081017.8, called 'Hamilton's Object' \citep{bib:Griffiths2021} and several years later dubbed 'The Scream' or 'The Cry' by \citep{bib:Ebeling2025}, is a peculiar example of such a highly detailed multiple-image configuration in a recently discovered galaxy cluster, RM J223013.1-080853.1 \citep{bib:Rykoff2016} or eMACSJ2229.9-0808 \citep{bib:Ebeling2025}.  
This galaxy cluster is not as rich in multiple-image configurations as others, for instance, the Hubble Frontier Field clusters \citep{bib:Lotz2017} (see \cite{bib:Richard2014} for an overview of multiple-image configurations), nor does the multiple-image configuration cover a significant region in the cluster centre such as the almost Einstein ring in Abell~3827. 

Apart from the highly detailed triple, \cite{bib:Griffiths2021} discovered a faint arc (see their Fig.~5) and \cite{bib:Ebeling2025} reported another triple-image configuration (see their Fig.~B93). 
Setting up lens models based on such a sparse data set is highly degenerate, and a global mass density profile is not well constrained.
Yet, it was possible to extract the local lens properties from the highly detailed image-triple based on seven features.
These features could be clearly identified within each multiple image and could also be matched across all three images, as detailed in \cite{bib:Griffiths2021} and \cite{bib:Lin2022}. 
An analysis of the local reduced shear directions revealed that the lensing galaxy cluster is most probably a merger, which was confirmed by \cite{bib:Ebeling2025}: spectroscopic redshifts of 117 galaxies covering the area of the galaxy cluster suggested at least two merging clumps, one around redshift $z_\mathrm{d1}\approx0.595$ and one around $z_\mathrm{d2}\approx0.625$ (see \cite{bib:Ebeling2025}, Fig.~B93). 
Hence, the cluster redshift previously determined to be $z_\mathrm{d}=0.526\pm0.018$ by photometric redshifts from SDSS was updated.

Concerning the redshift of the highly detailed triple-image configuration, a discrepancy between \cite{bib:Griffiths2021} and \cite{bib:Ebeling2025} arises. 
\cite{bib:Griffiths2021} used the Blue Keck Cosmic Web Imager (Blue KCWI) integrated-field unit and obtained $z_\mathrm{s}=0.8200\pm 0.0005$ as the redshift of the lensed source, while \cite{bib:Ebeling2025} employed $K$-band spectra obtained with Subaru/MOIRCS  in the infrared and arrived at $z_\mathrm{s}=3.201$, without determining an uncertainty.

The analysis of \cite{bib:Griffiths2021} included a comparison of the spectra of all three multiple images with each other (see their Fig.~9). 
These observations also covered a star-forming region (SFR) in the outskirts of the source galaxy in all three images.
\cite{bib:Griffiths2021} identified over ten absorption and emission features in the spectra of all three multiple images extending out to the SFR in the outskirts and arrived at $z_\mathrm{s} \approx 0.82$ (see their Table~3). 
Most of these lines were iron or magnesium lines, with one titanium, so that it remained questionable whether such a spectrum was realistic for the lensed galaxy. 
From Subaru/MOIRCS data, \cite{bib:Ebeling2025} identified the oxygen doublet $\left[\text{OIII}\right]\lambda\lambda4959,5007$ for the innermost multiple image in the fold configuration, image~A in \cite{bib:Griffiths2021}.
The presence of this (single) emission line doublet was deemed a unique indicator for $z_\mathrm{s}=3.201$. 

To resolve this ambiguity, we reanalyse the Blue KCWI spectra in this work; we organise the paper as follows:
Section~\ref{sec:kcwi} contains a new analysis of the Blue KCWI observations because the determination of a galaxy redshift with $z\approx1$ and $z\approx3$ is known to be difficult and the original evaluation of \cite{bib:Griffiths2021} did not consider $z\approx3$. 
Depending on the resolution and signal-to-noise ratio of the instrument, different emission and absorption lines can appear at similar observed wavelengths. 
Section~\ref{sec:moircs} then contains the details of the Subaru/MOIRCS observations, which show that an analysis of the data available from \cite{bib:Ebeling2025} is insufficient to constrain the redshift of the source galaxy with high certainty. 
Section~\ref{sec:conclusions} summarises our findings and presents our conclusions.


\begin{figure*}
\centering
\includegraphics[width=0.18\linewidth]{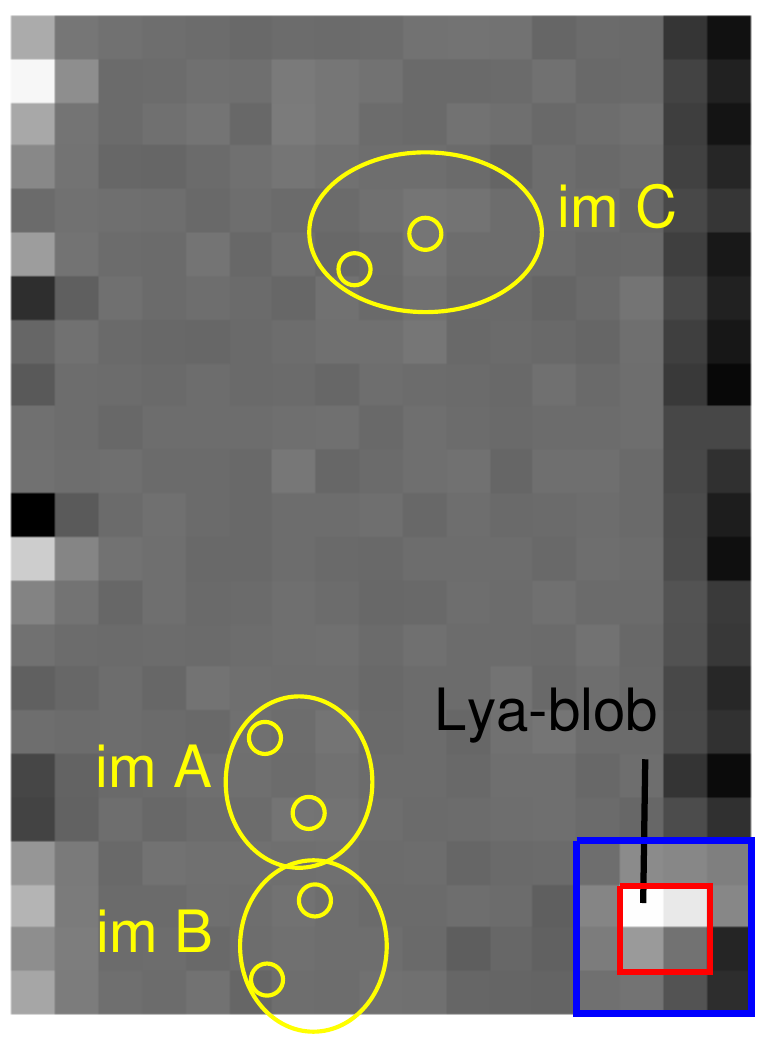}
\includegraphics[width=0.325\linewidth]{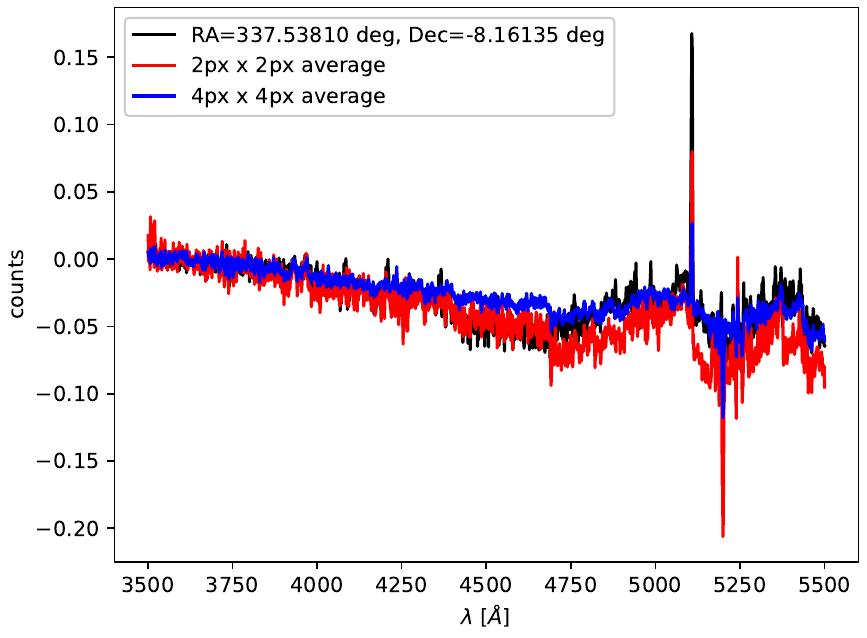}
\includegraphics[width=0.45\linewidth]{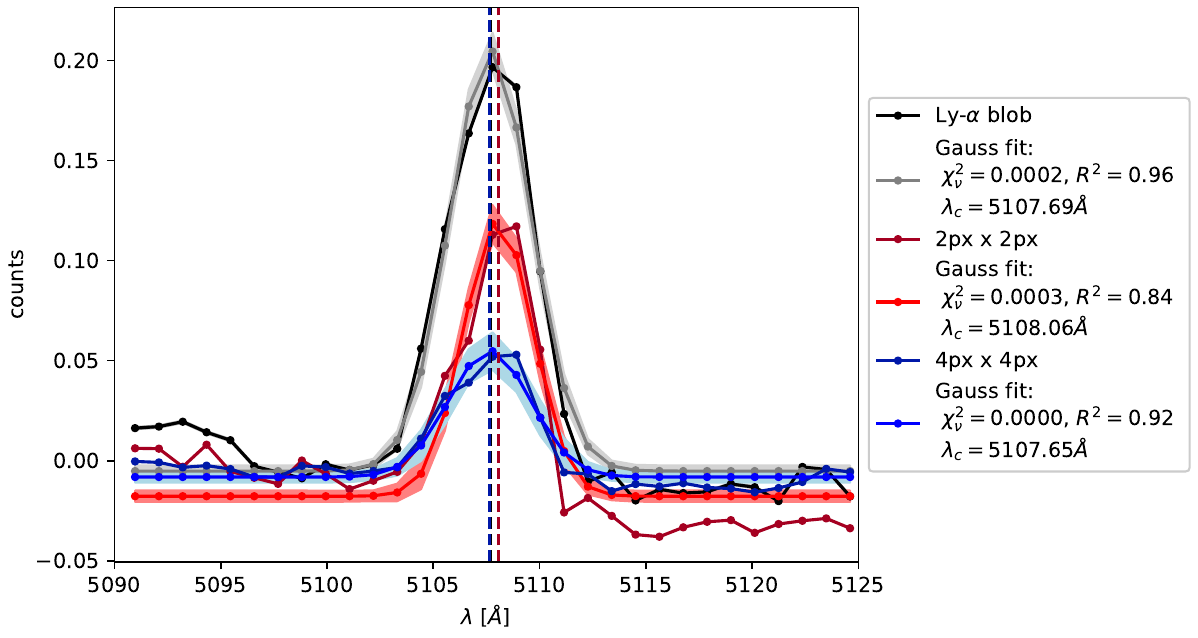}
\caption{Left: Slice of the KCWI data cube in which the Ly-$\alpha$ blob has its peak flux (white pixel). The red and blue squares mark the 2~px~$\times$~2~px and 4~px~$\times$~4~px regions around the Ly-$\alpha$ blob whose averaged raw spectra are shown in the central plot in the same colour. Centre: Raw spectra of the Ly-$\alpha$ blob based on the single pixel (black), the 2~px~$\times$~2~px (red), and the 4~px~$\times$~4~px region (blue) marked in the left image. Right: Zoom into the wavelength range around the Ly-$\alpha$ blob: background-subtracted spectra (dark lines, black, red, blue), error bars from the \texttt{PypeIt} reduction are plotted around the single-pixel spectrum, but are not visible. The grey, bright red, and bright blue lines including the pale-shaded areas show a Gaussian fit with 1-$\sigma$ confidence bounds. The yellow regions in the left plot show the three multiple images as identified in Section~\ref{sec:kcwi_multiple-image_identification} to illustrate that cross-contamination of their spectra by the Ly-$\alpha$ blob is small.}
\label{fig:lya}
\end{figure*}

\section{Blue KCWI redshift determination}
\label{sec:kcwi}

As detailed in Appendix~\ref{sec:data_reduction}, the data used in this section was taken with the KCWI on September 16, 2020 under the pro-
gramme ID H315 with Richard E. Griffiths as Principal Investigator. 
Details on the data acquisition specifics can be found in the appendix in Section~\ref{sec:kcwi_data_acquisition}. 
Our completely new data calibration and reduction of the existing dataset is outlined in the appendix in Sections~\ref{sec:kcwi_data_calibration} and \ref{sec:kcwi_data_cube}. 
In the following, the newly created data cube is used to mainly reinvestigate the redshift of the background galaxy of the triple-image configuration. 
But prior to this, as a clean feasibility test for the spectrum extraction, we \comm{first} reproduced the spectrum for the Lyman-$\alpha$ blob discovered in \cite{bib:Griffiths2021}.

\subsection{Lyman-\texorpdfstring{$\alpha$}{a} blob spectral analysis}
\label{sec:kcwi_lya_spectral}

The Ly-$\alpha$ blob, first detected in \cite{bib:Griffiths2021}, is a distinct background object at $z\approx3.20$.
It is unrelated to Hamilton's Object physically and not linked to the multiple images of its bulge or to its SFR. 
The Ly-$\alpha$ blob is located along a different line of sight and only serendipitously appears in projection within the same KCWI field of view.
Therefore, its significance in our analysis is purely methodological: it serves as an independent, spatially isolated single emission-line source at a well-determined redshift, which we use to validate our data reduction pipeline and spectrum extraction procedure before applying the same methods to the more complex triple-image configuration.

The KCWI data cube was read in and all 49 bad pixels, indicated by the bad pixel mask, were marked as \texttt{NaN} to be excluded from any analysis.
Subsequently, the strong Lyman-$\alpha$ emitter was traced in the data cube to be at RA$=337.53810$~deg and Dec$=-8.16135$~deg, see Fig.~\ref{fig:lya} (left). 
This position is only 2.5~arcsec from that found in \cite{bib:Griffiths2021}, which is located one pixel to the right and one pixel down from our position. 
As Fig.~\ref{fig:lya} (left) shows the slice of the KCWI data cube at which the Ly-$\alpha$ blob has its peak emission, it is clear from the plot that the blob does not have a second image above the detection limit close to image~B. 
It is thus not multiply-imaged in the same configuration as Hamilton's Object within the KCWI field of view.

To study its potentially contaminating impact on the multiple images (marked for comparison in yellow in Fig.~\ref{fig:lya} (left)), we plotted \comm{the Ly-$\alpha$ blob's} single-pixel spectrum (black line) and the spectrum that is obtained as the average spectrum from two regions of different sizes including the Ly-$\alpha$ blob.
The pixel masks to extract the three spectra around the Ly-$\alpha$ blob are indicated by the red and blue squares in Fig.~\ref{fig:lya} (left), the spectra themselves are shown in Fig.~\ref{fig:lya} (centre) in corresponding colours. 
From the decreasing peak amplitude of the three spectra in Fig.~\ref{fig:lya} (centre), it is thus clear that the blob does not cause any significant contamination of the image triple. 
Moreover, in agreement with \cite{bib:Griffiths2021}, the extent of the blob does not exceed 5.4~arcsec (4~px~$\times$~4~px). 

To determine the redshift from the peak position along the spectral axis, we used the single-pixel spectrum and also plotted the error bars as saved in the 'SIG' data cube as standard deviations around the wavelength-count pairs. 
Since these error bars are small, they are not visible in Fig.~\ref{fig:lya} (right). 
Then we assumed that the background is the median of the pixel's spectrum excluding $\lambda \in \left[5090, 5125 \right]$~\AA  \ and subtracted this value from the entire spectrum. 
Afterwards a Gaussian with overall offset was fitted to the spectral line within $\lambda \in \left[5090, 5125 \right]$~\AA \ employing  \texttt{LMFit} \cite{bib:Newville2025}. 
The result is shown in Fig.~\ref{fig:lya} (right) as a grey line with confidence bounds of 1-$\sigma$ around the fit. 
We read off the peak position at $\lambda_\mathrm{obs}=(5107.69\pm 1.12)$~\AA, with the uncertainty given by the wavelength calibration because the one from the fit was one order of magnitude smaller.
Repeating the procedure of background subtraction and Gauss fitting for the extended pixel masks shown in Fig.~\ref{fig:lya} (left), $\lambda_\mathrm{obs}\approx 5108$\AA \ remains a robust result, yet, at a lower quality of fit, as indicated by $\chi_\nu^2$ and $R^2$.
Consequently, as obtained from the single-pixel spectrum with the highest quality of fit, 
\begin{equation}
z_\mathrm{c} = \frac{\lambda_\mathrm{obs}}{\lambda_\mathrm{Ly\alpha}} - 1 = 3.204 \;, \quad \Delta z_\mathrm{c} = \frac{\Delta \lambda_\mathrm{obs}}{\lambda_\mathrm{Ly\alpha}} = 0.001 \;.
\label{eq:lya}
\end{equation}
This result disagrees with $z_\mathrm{c}=3.199\pm0.001$ from \cite{bib:Griffiths2021} within the confidence bounds. 
However, it is close enough, considering that \cite{bib:Griffiths2021} arrived at their value by manual evaluation and given the fact that the entire peak is only formed by about ten data points.

\begin{figure*}
\centering
\includegraphics[width=0.32\linewidth]{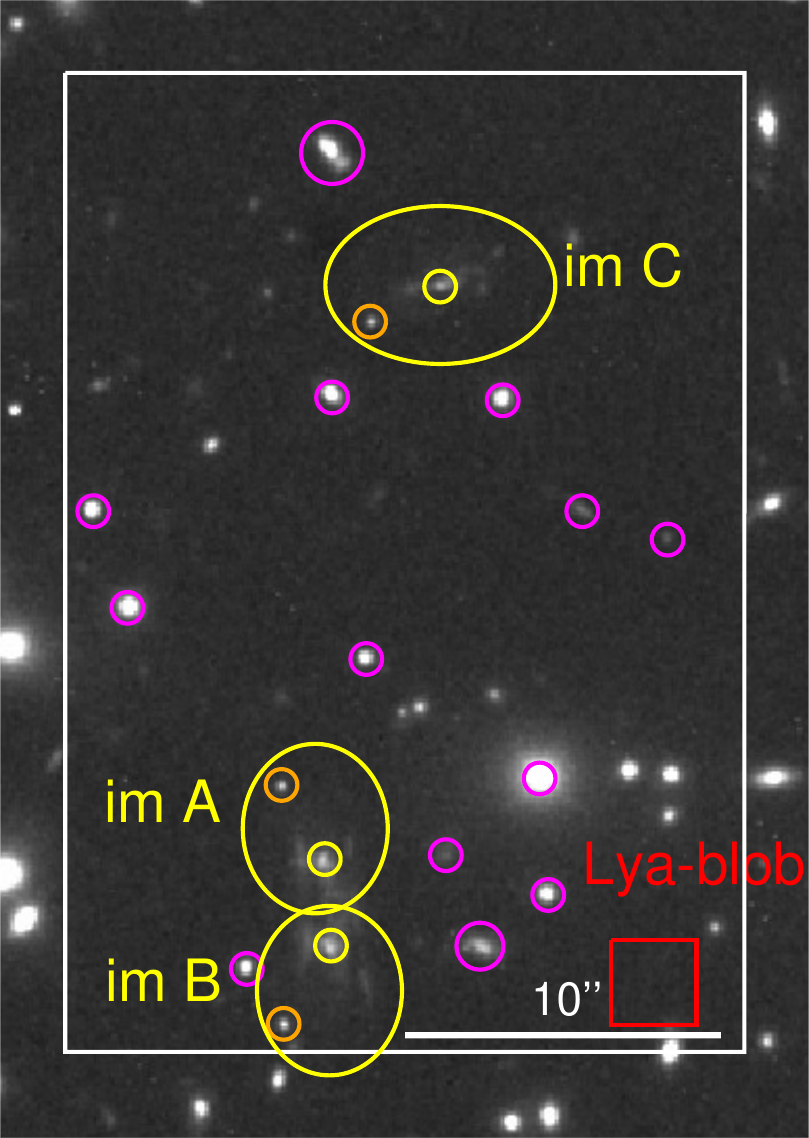}
\raisebox{2ex}{\includegraphics[width=0.29\linewidth]{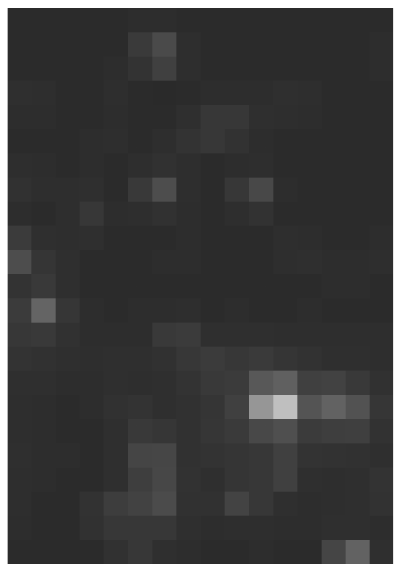}}
\raisebox{2ex}{\includegraphics[width=0.295\linewidth]{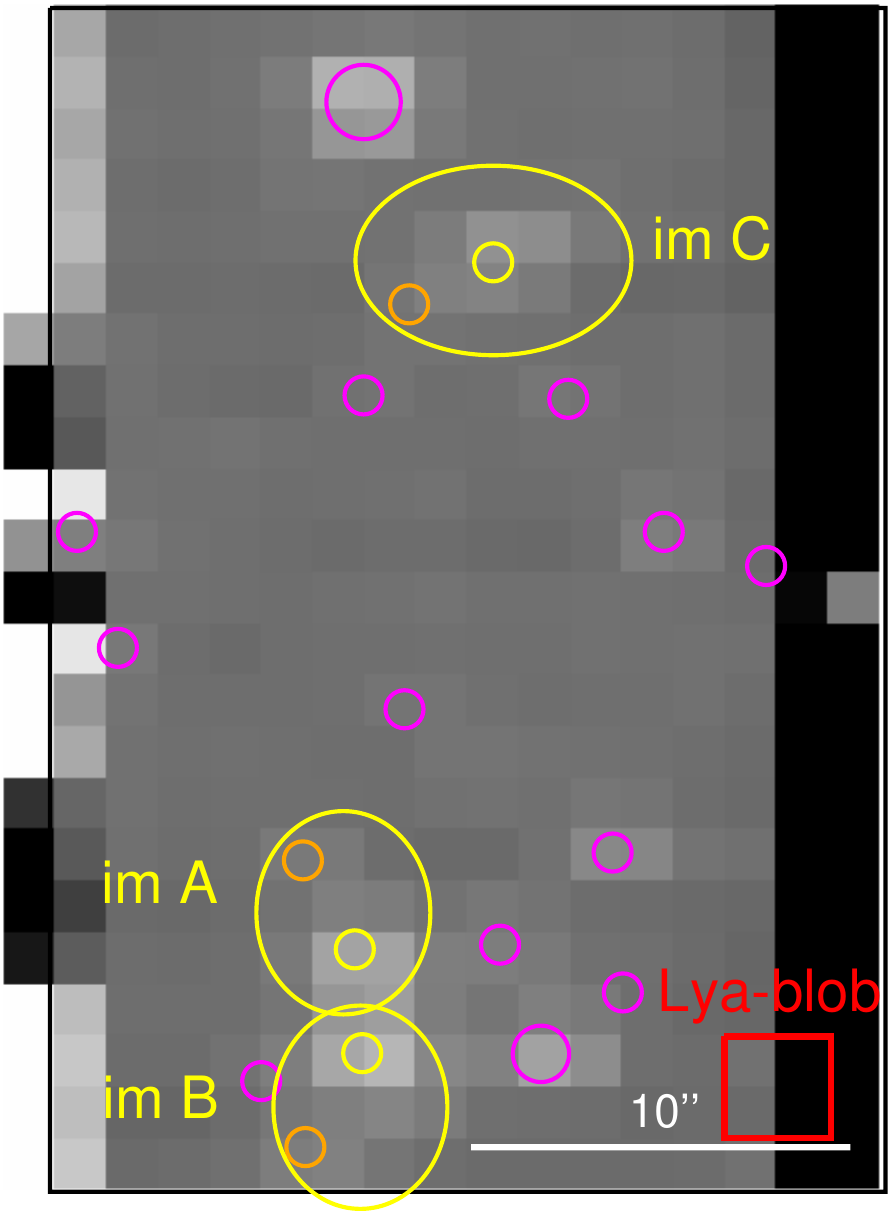}}
\caption{Left: HST F110W filter-band observation with multiple images marked by yellow ellipses. The bulge and prominent star-formation region in the outer part of the galaxy are marked by yellow \comm{and orange} circles, \comm{respectively}. Magenta circles highlight additional bright objects in the field that may leave signatures in the white-light image of the KCWI data cube. Centre: Downsampled version of the HST F110W filter-band observation (in the white rectangle) to match the resolution of the KCWI white-light image. Right: KCWI white-light image with yellow-magenta regions overlaid such that the regions coincide with the brightness maxima.
The spectra were extracted from the KCWI data cube integrated over all pixels inside the yellow pixel masks (large ovals and small circles). For reference, we also include the position of the Ly-$\alpha$ blob, marked by the red box, discussed in Section~\ref{sec:kcwi_lya_spectral}.}
\label{fig:cube_fit1}
\end{figure*}

\subsection{Multiple-image identification}
\label{sec:kcwi_multiple-image_identification}

To extract the spectra for the multiple images, the corresponding sky-pixels in the data cube need to be identified. 
From the \textit{Hubble} Space Telescope (HST) WFC3 UVIS F814W filter-band observation and WFC3 IR F110W, we set up masks for all three multiple images: these masks cover the entire image area of each multiple image and additionally mark the central bulge region and the SFR in the outer part of the galaxy separately; see Fig.~\ref{fig:cube_fit1} (left, yellow ellipses and circles).
Even though the world coordinate system (WCS) should be aligned for both HST observations, there is an offset between the two.
Similarly, an offset is also visible when these masks are laid over the white-light image of the KCWI data cube. 
To find optimum alignment, we marked some bright galaxies in the environment of the triple- image configuration (see magenta circles in Fig.~\ref{fig:cube_fit1}, left) and shifted the pixel mask so that the yellow and magenta masks matched the brightest pixels in the white-light image of the KCWI data cube. 
The result is shown in Fig.~\ref{fig:cube_fit1} (right) and is based on the assumption that bright galaxies also have a significant brightness in the white-light image, implying that they have strong emission characteristics in the optical wavelength bands. 
Applying this translation to the Ly-$\alpha$ blob position from \cite{bib:Griffiths2021}, it coincides with our position, which corroborates its necessity. 

The HST WFC3 IR F110W observation has a spatial conversion factor of 0.12825~arcsec per pixel, a higher resolution than the 1.35~arcsec per spatial pixel of the KCWI data cube. 
Thus, to create pixel masks on the sky from which spectra along the third dimension of the data cube can be obtained, we extracted the centres and extensions of the yellow masks overlaid over the KCWI data cube.
Fig.~\ref{fig:cube_fit1} shows the result. 
As the coordinates had to be rounded to integers, the previously circular masks for the bulge and SFR were only 1 pixel wide.
There is a slight imbalance in the number of pixels per mask for the entire image area: while the ones for images~A and B cover 6~pixels, the one for image~C covers 8~pixels. 

\begin{figure*}
\centering
\includegraphics[width=0.49\linewidth]{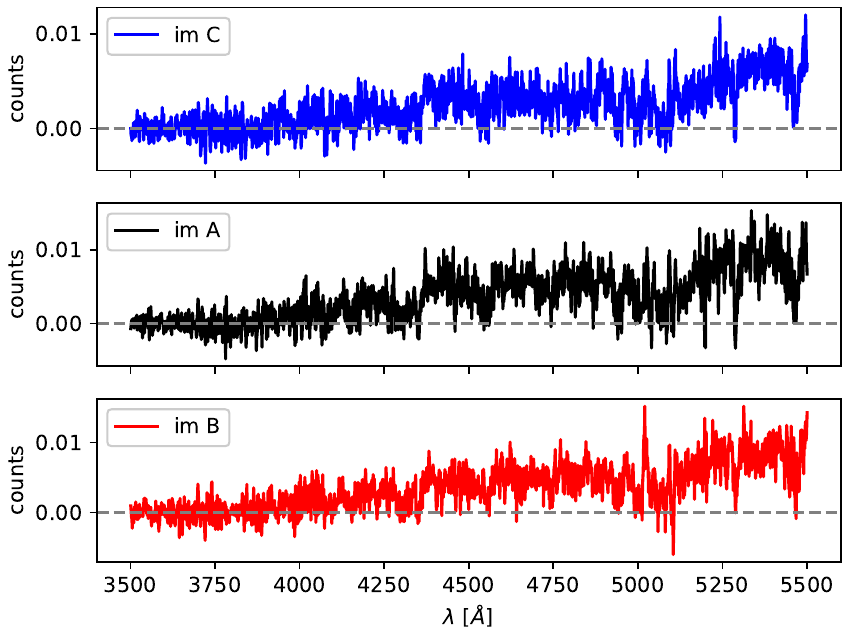} \hfill
\includegraphics[width=0.49\linewidth]{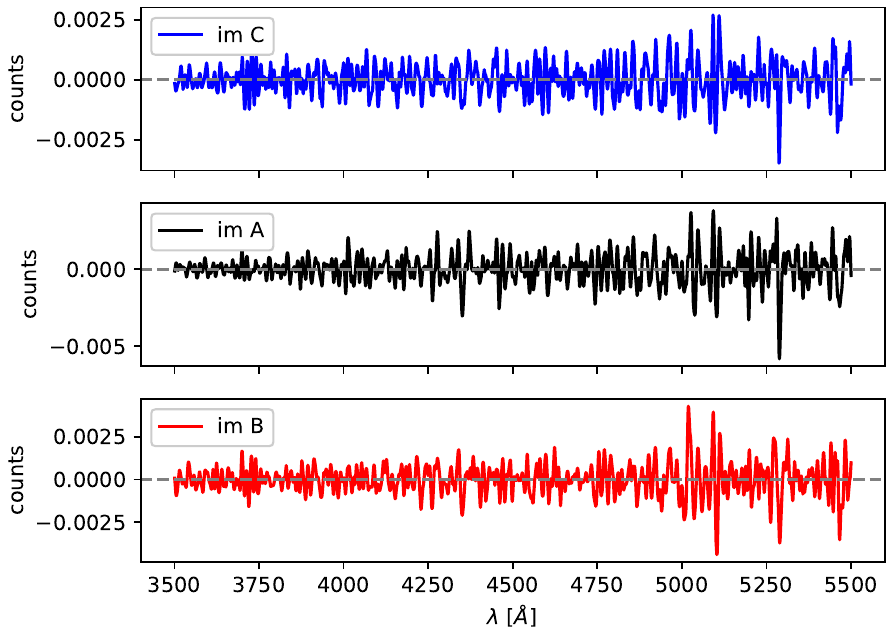}
\caption{Left: Raw spectra extracted from the total image areas. The order from top to bottom is chosen to be the same as in the observation. Right: Background and continuum subtracted spectra which were smoothed with a Gaussian filter having $\sigma=2$~\AA.}
\label{fig:mis_spectra}
\end{figure*}

\subsection{Background and continuum subtraction}
\label{sec:kcwi_background_subtraction}

Similarly to the evaluation of the Ly-$\alpha$ blob, we first plotted the raw spectra for all nine pixel masks of the multiple images. 
The single-pixel spectra of the bulges and SFRs were quite noisy, so Fig.~\ref{fig:mis_spectra} (left) only shows the spectra extracted from the ellipses of the total image areas. 
Due to the unknown locations of the absorption and emission lines, the background estimation could not be performed in the same way as for the single emission-line spectrum of the Lyman-$\alpha$ blob. 
Moreover, Fig.~\ref{fig:mis_spectra} (left) shows that the spectra of all three multiple images have a continuum emission underneath the individual absorption and emission lines.
Without fluxing, there is no information to be extracted from the continuum, which is why it was subtracted with the background for our redshift determination. 
We estimated the local background plus the continuum of the multiple-image spectra by convolving each spectrum with a broad Gaussian filter of $\sigma=10$~\AA, which means 9 pixels with 1.12~\AA \ per pixel.
Subtracting this highly smoothed version from each raw spectrum, we obtained the line spectra for the three multiple images. 
The width of the filter was chosen to be twice the full width at half maximum of the Ly-$\alpha$ blob, to smooth out the broad spectral features. 
To account for aliasing effects and noise in data acquisition, we convolved these spectra with a narrow Gaussian filter of $\sigma=2$~\AA, which means 2 pixels with 1.12~\AA. 
Although this decreases the resolution, it simplifies the data evaluation, as shown in Fig.~\ref{fig:mis_spectra} (right).
Comparing the three resulting spectra in Fig.~\ref{fig:mis_spectra} (right) with each other, it is obvious that they are highly correlated and actually belong to multiple images from the same source.

\subsection{Line identification}
\label{sec:kcwi_line_identification}

 \begin{figure*}
\sidecaption
\includegraphics[width=12cm]{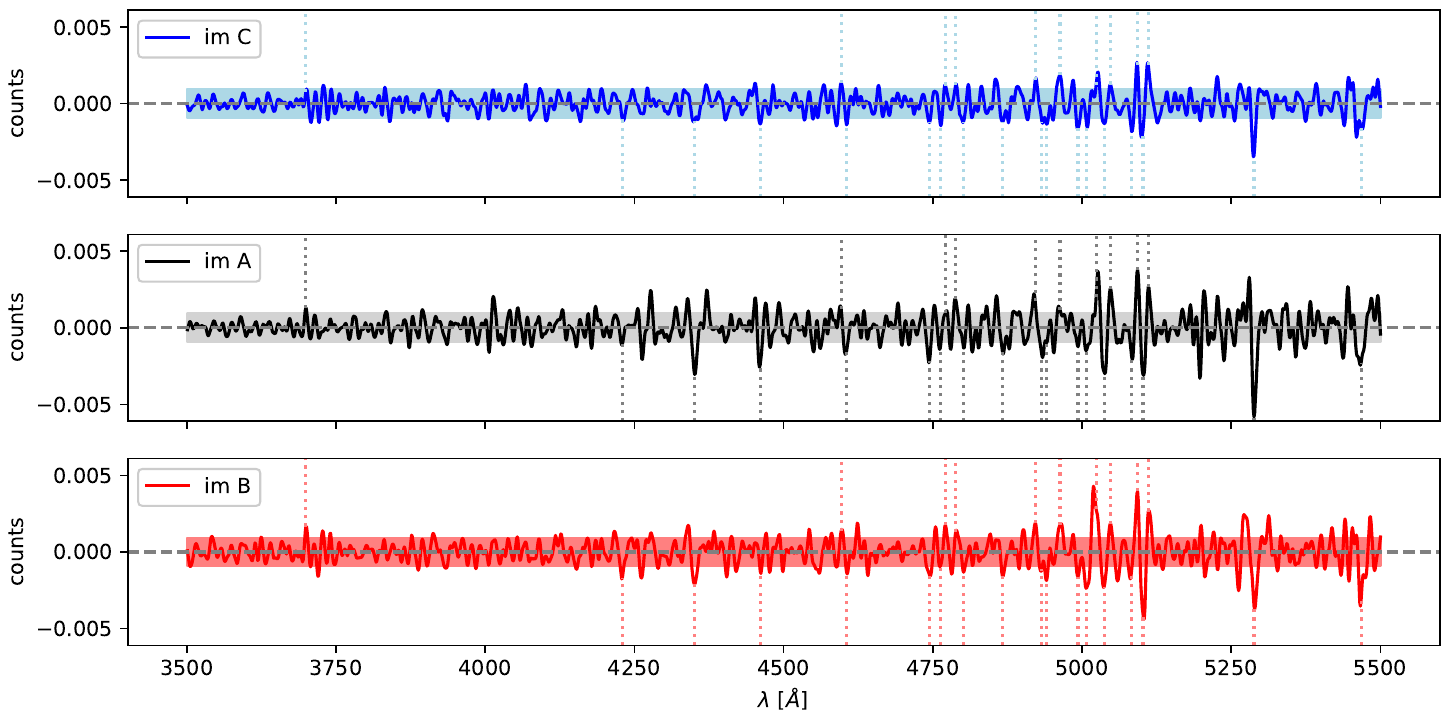}
\caption{Background and continuum-subtracted spectra of all three multiple images, smoothed with a Gaussian filter of 2~\AA \ width to alleviate aliasing, same as Fig.~\ref{fig:mis_spectra} (right). The 3-$\sigma$ band around zero based on the background noise level is shaded. The 17 absorption and ten emission lines above 3 $\sigma$ are marked in each spectrum as dotted lines upwards (emission) and downwards (absorption), respectively.}
\label{fig:mis_spectra_final}
\end{figure*}

To extract significant absorption and emission lines from Fig.~\ref{fig:mis_spectra} (right), the definition of a noise level is required.
We chose three pixel masks at 6 pixels each on a side in the vicinity of the three multiple images, which did not show any strong spectral features in a visual inspection.
Subsequently, we calculated the median spectrum of all 18 pixels jointly and smoothed the latter with a broad Gauss filter of $\sigma=10$~\AA. 
The square root of the variance of this smoothed background spectrum yields a noise level estimate of $\sigma=0.00031$, which we used as a 1-$\sigma$ confidence bound around zero for the spectra in Fig.~\ref{fig:mis_spectra} (right).
Determining the noise level for the individual pixel masks close to all three multiple images~A, B, and C, we arrived at $\sigma=0.00037, 0.00035$, and $0.00044$, respectively.
The increased values compared to those from the joint spectrum originate from the reduced number of pixels. 
The noise level as determined from the actual pixel masks in all three multiple images amounts to $\sigma=0.00021, 0.00020$, and $0.00015$. 
Thus, using the larger value of the joint spectrum, we chose a conservative bound on the noise level.  

Extracting all absorption and emission lines with an amplitude of at least 3-$\sigma$ above zero in the spectrum of each multiple image with the \texttt{scipy} routine \texttt{find\_peaks}, there are more than 38 emission lines and more than 42 absorption lines detected in each spectrum. 
Only considering those that are common to all three spectra within 2~\AA, we reduced the number to 16 absorption and nine emission peaks, as shown in Fig.~\ref{fig:mis_spectra_final}. 
The 16 absorption lines are at observed wavelengths $\lambda_\mathrm{obs}$ of 4230~\AA, 4351~\AA, 4461~\AA, 4605~\AA, 4745~\AA, 4762~\AA, 4802~\AA, 4867~\AA, 4933~\AA, 4941~\AA, 4993~\AA, 5008~\AA, 5037~\AA, 5083~\AA, 5288~\AA, 5468~\AA. 
The nine emission lines having $\lambda_\mathrm{obs}$ read 3700~\AA, 4597~\AA, 4771~\AA, 4788~\AA, 4922~\AA, 4963~\AA, 5047~\AA, 5093~\AA, 5111~\AA.
Moreover, since the two peaks are broader and are not included in this selection, we also included the absorption feature at an average observed wavelength of 5102~\AA \ and the emission line at an average observed wavelength of 5024~\AA \ in the set of identified lines common to all three spectra.

\begin{table*}[t]
\caption{Observed spectral features above 3 $\sigma$ and possible identifications as emission and absorption lines.}
\centering
\begin{tabular}{cc|cccc|cccc}
 \hline \hline
$\lambda_\mathrm{obs}$ &	SNR	&	line	&	$\lambda_\mathrm{em}$	&	$z_{0.82}$	&	$\Delta z_{0.82}$	&	line	&	$\lambda_\mathrm{em}$	&	$z_{3.20}$	&	$\Delta z_{3.20}$	\\
$\left[ \text{\AA} \right]$ &  & ($z\approx 0.82$) & $\left[ \text{\AA} \right]$ & & & ($z\approx 3.20$) & $\left[ \text{\AA} \right]$ & & \\
\hline																			
4230	&	3.6	&		&		&		&		&		&		&		&		\\
4351	&	5.8	&	\textit{FeII}	&	2396	&	0.816	&	0.002	&	\textit{OIV}	&	1034	&	3.208	&	0.004	\\
4461	&	5.6	&		&		&		&		&		&		&		&		\\
4605	&	4.8	&		&		&		&		&		&		&		&		\\
4745	&	5.1	&	MnII	&	2606	&	0.821	&	0.002	&		&		&		&		\\
4762	&	4.6	&	FeII	&	2618	&	0.819	&	0.002	&		&		&		&		\\
4802	&	4	&		&		&		&		&		&		&		&		\\
4867	&	4.4	&		&		&		&		&		&		&		&		\\
4933	&	4.8	&		&		&		&		&		&		&		&		\\
4941	&	3.9	&	\textit{FeI}	&	2719	&	0.817	&	0.001	&	CIII	&	1176	&	3.202	&	0.003	\\
4993	&	4.8	&		&		&		&		&		&		&		&		\\
5008	&	4.9	&		&		&		&		&	SiII	&	1192	&	3.201	&	0.003	\\
5037	&	6	&		&		&		&		&	NI	&	1200	&	3.198	&	0.003	\\
5083	&	6.1	&	MgII	&	2796	&	0.818	&	0.001	&		&	&		&	\\
5102	 &	10.4	&	MgII	&	2803	&	0.820	&	0.001	&	Ly-$\alpha$	&	1215	&	3.199	& 0.003		\\
5288	&	13.6	&	FeII	&	2905	&	0.820	&	0.001	&	SiII	&	1260	&	3.197	&	0.003	\\
5468	&	6.2	&	NiI	&	3004	&	0.820	&	0.001	&	OI	&	1302	&	3.200	&	0.003	\\
\hline																			
3700	&	3.4	&		&		&		&		&		&		&		&		\\
4597	&	4.4	&		&		&		&		&		&		&		&		\\
4771	&	3.8	&	FeII*	&	2626	&	0.817	&	0.002	&		&		&		&		\\
4788	&	4.8	&	FeII*	&	2632	&	0.819	&	0.002	&		&		&		&		\\
4922	&	6	&		&		&		&		&		&		&		&		\\
4963	&	5.5	&		&		&		&		&		&		&		&		\\
5024	&	$>3$	&		&		&		&		&		&		&		&		\\
5047	&	5.4	&		&		&		&		&		&		&		&		\\
5093	&	9.8	&	MgII	&	2796	&	0.822	&	0.001	&	\multirow{2}{*}{Ly-$\alpha$} 	&	\multirow{2}{*}{1215}	&	\multirow{2}{*}{3.199}	&	\multirow{2}{*}{0.003}	\\
5111	&	8.6	&	MgII	&	2803	&	0.823	&	0.001	&		&		&		&	\\
\hline																			
\end{tabular}
\tablefoot{Columns: Observed wavelength, $\lambda_\mathrm{obs}$, S/N, name of the line, rest-frame emission wavelength, $\lambda_\mathrm{em}$, estimated redshift and uncertainty, analogous to Eq.~\eqref{eq:lya}. The analysis is performed for lines around $z\approx0.82$ (columns 3-6) and $z\approx3.20$ (columns 7-10). The upper part lists absorption lines, the lower one emission lines. Italic printed line names are suboptimal identifications and not taken into account in the overall redshift determination.
\\
Extending the wavelength range beyond 5500~\AA, \cite{bib:Griffiths2021} identified an absorption feature at 5600~\AA \ attributed to an \comm{observationally rare} TiII line with $\lambda_\mathrm{em}=3075$~\AA \ at $z=0.82$. For $z\approx 3.2$, this line is a more common CII line with $\lambda_\mathrm{em}=1334.5$~\AA.}
\label{tab:lines}
\end{table*}

The maximum uncertainty in the observed wavelengths is 4~\AA \ except for the broad emission peak around 5024~\AA \ for which we assume 6~\AA. 
These values are adopted as conservative error bars in the following. 
As these 27 spectral features are common to all three multiple-image spectra, the probability that they were caused by artefacts in the processing, for instance, the background subtraction, or by physical, biasing sources, such as cosmic rays, is very low. 
Eleven of these features also coincide with those already identified in Table~3 of \cite{bib:Griffiths2021}. 

\subsection{Triple-image redshift determination}
\label{sec:kcwi_redshift_determination}

As we already know from \cite{bib:Griffiths2021} and \cite{bib:Ebeling2025}, the expected redshift of the source could be $z=0.820$ or $z=3.201$. 
Each of these possible values was investigated, comparing the spectral features identified in Section~\ref{sec:kcwi_redshift_determination} to emission and absorption lines commonly observed in galaxies at these redshifts see, for instance, \cite{bib:Martin2012} and \cite{bib:Griffiths2021} for possible lines around $z=0.820$ and \cite{bib:Steidel2003}, \cite{bib:Shapley2006}, and \cite{bib:Verhamme2017} for possible lines around $z=3.201$ and peculiar double-peaked Ly-$\alpha$ emitters, as the background galaxy considered here could be one of them.

Table~\ref{tab:lines} summarises the resulting redshifts from the individual lines around $z\approx0.82$ and $z\approx 3.20$. 
For each spectral feature that was found and could be identified as an absorption or emission line, we individually determined the redshift. 
The final redshift was then obtained by the average of all clearly identified absorption or emission lines weighted by their S/N. 
As error bars, we calculated the square root of the weighted variance to determine the spread of individual redshift estimates around the mean. 
Based on six absorption lines each, we obtained these estimates:
\begin{equation}
z_\mathrm{abs} = 0.820 \pm 0.001 \;, \quad z_\mathrm{abs} = 3.199 \pm 0.002 \;.
\label{eq:zabs}
\end{equation}
Based on four emission lines for the lower redshift and one emission line for the higher redshift, we also obtained
\begin{equation}
z_\mathrm{em} = 0.821 \pm 0.002 \;, \quad z_\mathrm{em} = 3.199 \pm 0.003 \;.
\label{eq:zem}
\end{equation}
Thus, both redshift estimates are consistent when calculated from their emission and absorption lines and, within their confidence bounds, are in agreement with the original two estimates from \cite{bib:Griffiths2021} and \cite{bib:Ebeling2025}.

From this analysis, the lower estimate seems more likely because it is based on more lines with a smaller spread. 
In addition, there are several other factors that favour this lower estimate.
We detail these in the following.

 \begin{figure*}
\sidecaption
\includegraphics[width=12cm]{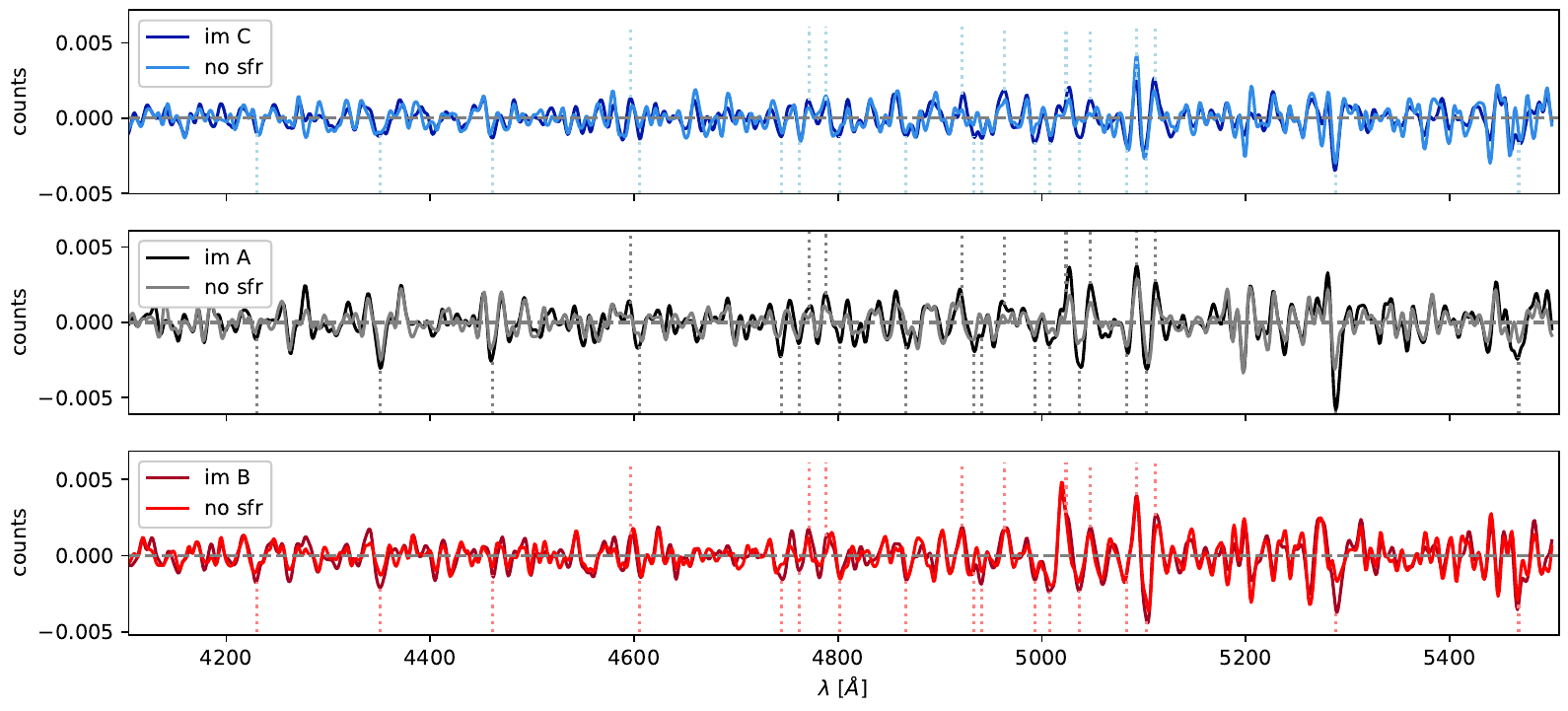}
\caption{Background and continuum subtracted spectra for all multiple images over the entire image regions (dark colours) as in Fig.~\ref{fig:mis_spectra_final} and omitting the SFR in the left part of the images (brighter colours). Dotted lines to the top and bottom mark emission and absorption lines above 3 $\sigma$ beyond noise level as listed in Table~\ref{tab:lines}, respectively. Salient regions are the lines around 4770~\AA \ as well as the last two significant absorption features.}
\label{fig:mis_spectra_nosfr}
\end{figure*}

First, the double peaks and local minima around 5100~\AA \ are more likely to be a resonant emission from a MgII doublet with an additional blueshifted absorption rather than a double Ly-$\alpha$ emission peak because according to \cite{bib:Feltre2018}, this Mg emission-absorption combination is quite common, while according to \cite{bib:Verhamme2017} double-peaked Ly-$\alpha$ emission is a rather peculiar feature of so-called Lyman-continuum emitters.
More recently, \cite{bib:Uzsoy2025} and references therein claim that double-peaked Ly-$\alpha$ emitters are more common, yet perhaps become rarer with increasing redshift. 
Moreover, \cite{bib:Verhamme2017} state that the emission peaks are typically about 300~km/s apart, implying a $\Delta z \approx 0.001$.
However, the peaks here are 18~\AA \ apart at 5102~\AA, so they yield $\Delta z \approx 0.0035$, 3.5 times larger than typical.
We also observe Ly-$\alpha$ in absorption, a very rare feature of Lyman continuum emitters, as stated in \cite{bib:Verhamme2017}.
Fig.~\ref{fig:mgii} (left) in \comm{Appendix~\ref{sec:additional_plots}} shows details of the spectra around 5100~\AA \ for further inspection. 
Comparing the relative height of the two emission peaks with Fig.~6 of \cite{bib:Uzsoy2025}, which shows a stacked double-peaked Ly-$\alpha$ profile for DESI/ODIN Ly-$\alpha$ emitters, we also observed that, on average, the blueshifted emission peak is much smaller than the redshifted one. 
This is also in stark contrast to the profile of our background source. 
 
Second, if we cut off the pixels that most likely contain the SFR in the outskirts, such that each multiple image is only covered by 4~px, we can plot the spectra without the SFR on top of the ones already shown in Fig.~\ref{fig:mis_spectra_final}. 
Fig.~\ref{fig:mis_spectra_nosfr} shows the result. 
For better visibility purposes, we cut the wavelength range at the left where there are no identified emission and absorption lines. 

As can be observed from the comparison of the spectra with and without the SFR, the most prominent features that lose amplitude are those around 4770~\AA \ and those at 5288~ \AA \ and 5468~\AA. 
For $z \approx 0.82$, these lines all belong to iron and the last also to nickel lines, which are typical for gas outflows in galaxies with SFRs \cite{bib:Martin2012}. 
While the FeII line at $\lambda_\mathrm{em}=2600$~\AA \ is observed below the 3-$\sigma$ significance threshold, a correlation between the three multiple images is observed to form an emission line around $\lambda_\mathrm{obs}=4732$~\AA. 
Similarly, if the FeII line at $\lambda_\mathrm{em}=2587$~\AA \ decays, emission lines are supposed to form at $\lambda_\mathrm{em}=2613$~\AA, $\lambda_\mathrm{em}=2632$~\AA \ with higher probability than resonant emission at $\lambda_\mathrm{em}=2587$~\AA.
The $\lambda_\mathrm{em}=2613$~\AA \ is observed around 4755~\AA \ in the spectra of images~A and B, but not significantly in C. 
All three show an emission line for $\lambda_\mathrm{em}=2632$~\AA. 
Hence, the iron line pattern seems to align and explain some spectral features in 4500-5100~\AA. 
Other spectral features observed in this wavelength range could also be attributed to absorption and emission lines of the galaxy cluster being a merger whose redshift falls between 0.526 and 0.62.
In contrast, assuming $z\approx3.20$ these spectral patterns either remain to be explained or have to be noise and no signatures of cluster emission or absorption are expected.
In Fig.~\ref{fig:4500-5100} in \comm{Appendix~\ref{sec:additional_plots}}, we plot the spectra of all multiple images for further details.

Finally, the fact that the 5468~\AA \ feature splits into further, finer lines can also be explained for $z\approx0.82$, as several iron and nickel lines would fall around this observed wavelength. 
In contrast, assuming $z\approx3.20$, the loss of amplitude and splitting into finer lines has to be explained by the absence of neutral oxygen and ionised silicon. 
Yet SiII has a longer rest-frame wavelength than OI, $\lambda_\mathrm{em}=1304$~\AA, and is thus at $\lambda_\mathrm{obs}=5478$~\AA, which would lead to $z_\mathrm{abs}=3.193\pm 0.003$. 
So it is harder to explain the last absorption feature and its change when excluding the SFR for $z\approx3.20$. 
Fig.~\ref{fig:mgii} (right) in \comm{Appendix~\ref{sec:additional_plots}} shows the details again.


\section{MOIRCS redshift determination}
\label{sec:moircs}

Subaru/MOIRCS observations of the multiple images were taken on September 20, 2019 under  programme ID 16321 with Harald Ebeling as Principal Investigator and retrieved from the SMOKA science archive\footnote{\url{https://smoka.nao.ac.jp/}}. 
They consist of 32 exposures of 180 seconds each.
The observations were taken at two telescope positions with a 1.5~arcsec offset for sky subtraction via the so-called ABBA dithering.
Each spatial pixel of the MOIRCS detector covers 0.116~arcsec on the sky, such that the seeing condition is the limiting factor for spatial resolution. 
As saved in the data files, the seeing ranged from 0.55~arcsec to 0.73~arcsec with an average of 0.64~arcsec. 
From Fig.~B93 of \cite{bib:Ebeling2025}, we read that the sky subtraction for image~A was performed along the long side of the slit, which had a position angle of 30~degrees.
Thus, two spectra with the same signals but different backgrounds can be extracted, as represented by a dark and a bright intensity blob to both sides of the SFR and the central bulge in image~A, corresponding to the 'B-A'  and the 'A-B' spectrum, respectively.
Based on this approach, it is unclear whether the sky subtraction for the bulge of image~A actually subtracted the background or the offset was still within the bulge of image~A, because the 1.5~arcsecond offset only corresponds to half the distance between the SFR and the centre of the bulge. 
Moreover, the galaxy also extends further to the other side away from the SFR as well.  
 
The grating used was HK500, optimised for 1.4-2.5~$\mu$m with R=500 for a slit width of 0.5~arcsec\footnote{\url{https://subarutelescope.org/Instruments/MOIRCS/spec_sensitivity.html}}. 
The spectral resolution is determined in the wavelength calibration. 
Although standard procedure suggests the use of an arc-lamp calibration for wavelengths below 2.5~$\mu$m, this calibration has to be done using OH nightsky line emissions because no arc-lamp calibration data were saved in the SMOKA science archive. 
Yet to resolve the OIII doublet with $\lambda_\mathrm{em} =4959$~\AA \ and $\lambda_\mathrm{em}=5007$~\AA \ at $z \approx 3.2$, we only need  $\Delta \lambda_\mathrm{obs} \approx 200$~\AA.
As can be observed in Fig.~B93 in \cite{bib:Ebeling2025}, this resolution is reached under these circumstances.
However, the noise level is quite inhomogeneous over the wavelength range close to 21,000~\AA \ around the redshifted OIII doublet. 
As a consequence, only the SFR of image~A yields a detection of the OIII doublet slightly above noise level and obeying the expected peak height ratio of 1:3 between the two peaks in the 'B-A' and 'A-B' spectra\footnote{Johan Richard, private communication.}. 
The 'B-A' and 'A-B' spectra for the bulge of image~A are inconclusive because they may be contaminated by the choice of the background, as mentioned above. 

Unfortunately, no slit was placed over image~B to collect data from this image.
The slit placed over image~C does not include the SFR and only covers the central bulge of the image, also missing the SFR.
Hence, a follow-up confirmation that the emission lines attributed to an OIII doublet in the SFR of image~A cannot be provided by the existing MOIRCS data. 

Given the high correlation between the spectra of all three multiple images in the Blue KCWI wavelength range, the MOIRCS data are much less coherent and lack the necessary observations.
Therefore, we refrained from performing a new analysis. 

As far as we know, the redshift of the Ly-$\alpha$ blob did not play a role in the assignment of the redshift of Hamilton's Object in \cite{bib:Ebeling2025}. As detailed in Sections~\ref{sec:kcwi_lya_spectral} and \ref{sec:conclusions}, the spatial coincidence of the blob with one candidate redshift is serendipitous and rather a hint against a high redshift of $z\approx3.2$.


\section{Conclusions}
\label{sec:conclusions}

In this work, we completely reanalysed the Blue KCWI data for three highly detailed multiple-lensed images within the galaxy cluster SDSS J223013 taken by Richard Griffiths on September 16, 2020 under the programme ID H315. 
We employed \texttt{PypeIt} to reduce the raw data into a data cube of two-dimensional spatial pixels and one-dimensional spectral pixels. 
Subsequent alignment and downsampling of HST observations allowed us to identify those spatial KCWI pixels that contain the three multiple images, including an SFR on the outskirts of the galaxy.
From all pixels belonging to a multiple image, we extracted its spectrum for $\lambda \in \left[3500,5500\right]$~\AA \ and identified absorption and emission peaks with an S/N of at least 3. 
The cut at 5500~\AA \ was necessary, as the spectra above this wavelength were found to be noisy and unstable. 
Therefore, the titanium line, observationally rather rare, that stems from a relatively low-abundance element with generally weak spectral features, around 5600~\AA \, as previously identified by \cite{bib:Griffiths2021}, could not be identified in our data cube again. 
An ordinary star-forming galaxy at redshift $z \approx 0.82$ would not usually be expected to show titanium in absorption at significant S/N.

Based on the emission and absorption features detected in the three multiple images, we listed those that were common to the spectra of all three multiple images and resulted in a redshift around 0.82 or 3.2 in Table~\ref{tab:lines}. 
We found that our best-fit value is $z_\mathrm{s}=0.8200\pm0.001$ from six absorption features and $z_\mathrm{s}=0.8201\pm0.002$ from four emission features. 
Further important spectral features supporting this lower redshift value were almost equal to the heights of the peaks around 5100~\AA \ because the interpretation as a double-peaked Lyman-$\alpha$ emission at the higher redshift was deemed \comm{physically atypical}, given the distance in wavelength and thereby redshift, the relative peak heights, and the fact that galaxies showing such a double-peaked emission are more rare than galaxies at lower redshift with a Mg II doublet, \cite{bib:Verhamme2017, bib:Uzsoy2025, bib:Feltre2018}. 
This result is also supported by the evaluation of the Gemini Multi-Object Spectrograph GMOS-N joint observation of both bulges of images~A and B in \cite{bib:Griffiths2021}: the OII doublet with $\lambda_\mathrm{em}=3726$~\AA \ and $\lambda_\mathrm{em}=3729$~\AA \, as well as the CaII H\&K break at $\lambda_\mathrm{em}=4000$~\AA  \,
and CN absorption at $\lambda_\mathrm{em}=4180$~\AA \ were identified. 
Together, these  features resulted in the first redshift estimate of the multiple-imaged galaxy to be $z=0.82$. 
Potentially, a CIII doublet emission with $\lambda_\mathrm{em}=1907$~\AA \ and $\lambda_\mathrm{em}=1909$~\AA \ around 8015~\AA \ could be identified as a hint for $z\approx3.2$. 
However, due to the low S/N and the decreased sensitivity of the GMOS-N around 5100~\AA \ and beyond 9700~\AA \, this spectrum could only give a first estimate and required follow-up observations with the Blue KCWI. 
Consequently, no emission or absorption features could be identified around 5100~\AA \ to distinguish the MgII doublet from a double-peaked Lyman-$\alpha$ emission and there were no emission and absorption features from oxygen lines beyond 9700~\AA \ to help resolve the ambiguity.  
 
To investigate the influence of the SFR, we excluded pixels that most likely belong to this part of the galaxy. 
Due to the low spatial resolution of 1.35~arcsec per spatial pixel, it was not possible to extract a separate spectrum for the SFR. 
Analysing the three multiple-image spectra without the SFR included, we found supporting consistency for the lower redshift because the resulting spectra mostly lost amplitude in iron and nickel lines, i.e.~lines that are actually expected in SFRs. 
Moreover, the reduction in amplitude for these lines also explained the more complex structure of the broad absorption peak around 5470~\AA.  
Putting the galaxy at the higher redshift, the more complex structure of this peak could not be explained. 

Unfortunately, the MOIRCS data taken by \cite{bib:Ebeling2025} were inconclusive upon follow-up investigation because the observations did not include image~B at all and also lacked the star-forming region of image~C. 
Moreover, we found that the ABBA dithering necessary for near-infrared observations requires more slits and careful positioning to ensure a clean background subtraction for the galaxy bulge. This does not seem to have been implemented in the observations in \cite{bib:Ebeling2025}.

Another complementary hint favouring $z_\mathrm{s}=0.82$ comes from the Ly-$\alpha$ blob. 
If we assume that the lensing potential does not change abruptly and the source of the Ly-$\alpha$ blob and the background galaxy being deflected into Hamilton's Object lie closely together in the source plane, the Ly-$\alpha$ blob should show a similar multiple-image configuration. Yet, within the KCWI field of view, the expected additional image of the Ly-$\alpha$ blob is missing.
While the hint favours $z_\mathrm{s}\approx0.82$, it is not a very strong one because the missing multiple images could lie outside of the KCWI field of view or the source of the blob could be sufficiently far away from the background galaxy.

Summarising the data evaluation in this work, it is important to note that multiple absorption and emission lines in the rest-frame ultraviolet wavelength range still cannot uniquely determine the redshift of a galaxy, even for most modern instruments such as the Blue KCWI.
Joining these results with spectra from complementary wavelength bands, particularly in the infrared, ambiguities can only be resolved with observations with spectral resolution R>2000. 

For the triple-image configuration in SDSS J223013, resolving the ambiguities will ascertain whether the configuration has the closest \comm{angular diameter} distance between the lensing galaxy cluster and the background source (490~Mpc for $z_\mathrm{d}=0.526$ and $z_\mathrm{s}=0.82$) that has been observed so far, at least for the lensing configurations listed in Table~\ref{tab:related_clusters}. 
One may argue that this unique proximity is the reason why Hamilton's Object is such a clear fold configuration without additional shear or slight rotations between the two images, as in the configuration of system 12 in MACSJ0416.
Resolving this ambiguity in the redshift is also important to constrain the mass of the lensing cluster when the observational data are combined with a model for the lensing mass density profile. 
Moreover, based on the previous estimates of the smoothness scale of dark matter in \cite{bib:Griffiths2021}, Hamilton's Object is also free of smaller-scale perturbations. 
Therefore, resolving the ambiguity in the redshift of Hamilton's Object will shed further light on the impact of line-of-sight perturbers on multiple images. 

\begin{acknowledgements}
We cordially thank the anonymous referee for very helpful and constructive comments to strengthen the paper.
This publication is based upon work from COST Action CA21136 Addressing observational tensions in cosmology with systematics and fundamental physics (CosmoVerse) supported by COST (European Cooperation in Science and Technology).
\\
Some of the data presented herein were obtained at Keck Observatory, which is a private 501(c)3 non-profit organization operated as a scientific partnership among the California Institute of Technology, the University of California, and the National Aeronautics and Space Administration. The Observatory was made possible by the generous financial support of the W. M. Keck Foundation.
The authors wish to recognize and acknowledge the very significant cultural role and reverence that the summit of Maunakea has always had within the Native Hawaiian community. We are most fortunate to have the opportunity to conduct observations from this mountain.
This research has made use of the Keck Observatory Archive (KOA), which is operated by the W. M. Keck Observatory and the NASA Exoplanet Science Institute (NExScI), under contract with the National Aeronautics and Space Administration.
\\
This research is based in part on data collected at the Subaru Telescope, which is operated by the National Astronomical Observatory of Japan. We are honored and grateful for the opportunity of observing the Universe from Maunakea, which has the cultural, historical, and natural significance in Hawaii.

\end{acknowledgements}

\bibliographystyle{aa}
\bibliography{ref}

@ARTICLE{bib:Bayliss2011,
       author = {{Bayliss}, Matthew B. and {Hennawi}, Joseph F. and {Gladders}, Michael D. and {Koester}, Benjamin P. and {Sharon}, Keren and {Dahle}, H{\r{a}}kon and {Oguri}, Masamune},
        title = "{Gemini/GMOS Spectroscopy of 26 Strong-lensing-selected Galaxy Cluster Cores}",
      journal = {\apjs},
         year = 2011,
        month = mar,
       volume = {193},
       number = {1},
          eid = {8},
        pages = {8},
          doi = {10.1088/0067-0049/193/1/8},
archivePrefix = {arXiv},
       eprint = {1010.2714},
 primaryClass = {astro-ph.CO},
       adsurl = {https://ui.adsabs.harvard.edu/abs/2011ApJS..193....8B}
}

@ARTICLE{bib:Caminha2017,
   author = {{Caminha}, G.~B. and {Grillo}, C. and {Rosati}, P. and {Balestra}, I. and 
	{Mercurio}, A. and {Vanzella}, E. and {Biviano}, A. and {Caputi}, K.~I. and 
	{Delgado-Correal}, C. and {Karman}, W. and {Lombardi}, M. and 
	{Meneghetti}, M. and {Sartoris}, B. and {Tozzi}, P.},
    title = "{A refined mass distribution of the cluster MACS J0416.1-2403 from a new large set of spectroscopic multiply lensed sources}",
  journal = {\aap},
archivePrefix = "arXiv",
     year = 2017,
   volume = 600,
      eid = {A90},
    pages = {A90},
      doi = {10.1051/0004-6361/201629297},
   adsurl = {http://cdsads.u-strasbg.fr/abs/2017A%26A...600A..90C}
}

@ARTICLE{bib:Diaz2021,
       author = {{D{\'\i}az-S{\'a}nchez}, A. and {Dannerbauer}, H. and {Sulzenauer}, N. and {Iglesias-Groth}, S. and {Rebolo}, R.},
        title = "{The Einstein Ring GAL-CLUS-022058s: a Lensed Ultrabright Submillimeter Galaxy at z = 1.4796}",
      journal = {\apj},
         year = 2021,
        month = sep,
       volume = {919},
       number = {1},
          eid = {48},
        pages = {48},
          doi = {10.3847/1538-4357/ac0f75},
archivePrefix = {arXiv},
       eprint = {2106.14281},
 primaryClass = {astro-ph.GA},
       adsurl = {https://ui.adsabs.harvard.edu/abs/2021ApJ...919...48D}
}

@ARTICLE{bib:Ebeling2025,
       author = {{Ebeling}, H. and {Richard}, J. and {Beauchesne}, B. and {Basto}, Q. and {Edge}, A.~C. and {Smail}, I.},
        title = "{Beyond MACS: physical properties of extremely X-ray luminous clusters at z > 0.5}",
      journal = {\mnras},
         year = 2025,
        month = mar,
       volume = {537},
       number = {3},
        pages = {2662-2694},
          doi = {10.1093/mnras/staf063},
archivePrefix = {arXiv},
       eprint = {2404.11659},
 primaryClass = {astro-ph.CO},
       adsurl = {https://ui.adsabs.harvard.edu/abs/2025MNRAS.537.2662E}
}

@ARTICLE{bib:Feltre2018,
       author = {{Feltre}, Anna and {Bacon}, Roland and {Tresse}, Laurence and {Finley}, Hayley and {Carton}, David and {Blaizot}, J{\'e}r{\'e}my and {Bouch{\'e}}, Nicolas and {Garel}, Thibault and {Inami}, Hanae and {Boogaard}, Leindert A. and {Brinchmann}, Jarle and {Charlot}, St{\'e}phane and {Chevallard}, Jacopo and {Contini}, Thierry and {Michel-Dansac}, Leo and {Mahler}, Guillaume and {Marino}, Raffaella A. and {Maseda}, Michael V. and {Richard}, Johan and {Schmidt}, Kasper B. and {Verhamme}, Anne},
        title = "{The MUSE Hubble Ultra Deep Field Survey. XII. Mg II emission and absorption in star-forming galaxies}",
      journal = {\aap},
         year = 2018,
        month = sep,
       volume = {617},
          eid = {A62},
        pages = {A62},
          doi = {10.1051/0004-6361/201833281},
archivePrefix = {arXiv},
       eprint = {1806.01864},
 primaryClass = {astro-ph.GA},
       adsurl = {https://ui.adsabs.harvard.edu/abs/2018A&A...617A..62F}
}

@ARTICLE{bib:Griffiths2021,
       author = {{Griffiths}, Richard E. and {Rudisel}, Mitchell and {Wagner}, Jenny and {Hamilton}, Timothy and {Huang}, Po-Chieh and {Villforth}, Carolin},
        title = "{Hamilton's Object - a clumpy galaxy straddling the gravitational caustic of a galaxy cluster: constraints on dark matter clumping}",
      journal = {\mnras},
         year = 2021,
        month = sep,
       volume = {506},
       number = {2},
        pages = {1595-1608},
          doi = {10.1093/mnras/stab1375},
archivePrefix = {arXiv},
       eprint = {2105.04562},
 primaryClass = {astro-ph.CO},
       adsurl = {https://ui.adsabs.harvard.edu/abs/2021MNRAS.506.1595G}
}

@ARTICLE{bib:Grillo2016,
       author = {{Grillo}, C. and {Karman}, W. and {Suyu}, S.~H. and {Rosati}, P. and {Balestra}, I. and {Mercurio}, A. and {Lombardi}, M. and {Treu}, T. and {Caminha}, G.~B. and {Halkola}, A. and {Rodney}, S.~A. and {Gavazzi}, R. and {Caputi}, K.~I.},
        title = "{The Story of Supernova {\textquotedblleft}Refsdal{\textquotedblright} Told by Muse}",
      journal = {\apj},
         year = 2016,
        month = may,
       volume = {822},
       number = {2},
          eid = {78},
        pages = {78},
          doi = {10.3847/0004-637X/822/2/78},
archivePrefix = {arXiv},
       eprint = {1511.04093},
 primaryClass = {astro-ph.GA},
       adsurl = {https://ui.adsabs.harvard.edu/abs/2016ApJ...822...78G}
}

@ARTICLE{bib:Jullo2007,
   author = {{Jullo}, E. and {Kneib}, J.-P. and {Limousin}, M. and {El{\'{\i}}asd{\'o}ttir}, {\'A}. and 
	{Marshall}, P.~J. and {Verdugo}, T.},
    title = "{A Bayesian approach to strong lensing modelling of galaxy clusters}",
  journal = {New Journal of Physics},
archivePrefix = "arXiv",
   eprint = {0706.0048},
     year = 2007,
    month = dec,
   volume = 9,
    pages = {447},
      doi = {10.1088/1367-2630/9/12/447},
   adsurl = {http://adsabs.harvard.edu/abs/2007NJPh....9..447J}
}

@ARTICLE{bib:Jullo2009,
   author = {{Jullo}, E. and {Kneib}, J.-P.},
    title = "{Multiscale cluster lens mass mapping - I. Strong lensing modelling}",
  journal = {\mnras},
archivePrefix = "arXiv",
   eprint = {0901.3792},
 primaryClass = "astro-ph.CO",
     year = 2009,
    month = may,
   volume = 395,
    pages = {1319-1332},
      doi = {10.1111/j.1365-2966.2009.14654.x},
   adsurl = {http://adsabs.harvard.edu/abs/2009MNRAS.395.1319J}
}

@ARTICLE{bib:Lagattuta2019,
       author = {{Lagattuta}, David J. and {Richard}, Johan and {Bauer}, Franz E. and {Cl{\'e}ment}, Benjamin and {Mahler}, Guillaume and {Soucail}, Genevi{\`e}ve and {Carton}, David and {Kneib}, Jean-Paul and {Laporte}, Nicolas and {Martinez}, Johany and {Patr{\'\i}cio}, Vera and {Payne}, Anna V. and {Pell{\'o}}, Roser and {Schmidt}, Kasper B. and {de la Vieuville}, Geoffroy},
        title = "{Probing 3D structure with a large MUSE mosaic: extending the mass model of Frontier Field Abell 370}",
      journal = {\mnras},
         year = 2019,
        month = may,
       volume = {485},
       number = {3},
        pages = {3738-3760},
          doi = {10.1093/mnras/stz620},
archivePrefix = {arXiv},
       eprint = {1904.02158},
 primaryClass = {astro-ph.GA},
       adsurl = {https://ui.adsabs.harvard.edu/abs/2019MNRAS.485.3738L}
}

@ARTICLE{bib:Lagattuta2023,
       author = {{Lagattuta}, David J. and {Richard}, Johan and {Ebeling}, Harald and {Basto}, Quentin and {Cerny}, Catherine and {Edge}, Alastair and {Jauzac}, Mathilde and {Mahler}, Guillaume and {Massey}, Richard},
        title = "{RXJ0437+00: constraining dark matter with exotic gravitational lenses}",
      journal = {\mnras},
         year = 2023,
        month = jun,
       volume = {522},
       number = {1},
        pages = {1091-1107},
          doi = {10.1093/mnras/stad803},
archivePrefix = {arXiv},
       eprint = {2303.09568},
 primaryClass = {astro-ph.CO},
       adsurl = {https://ui.adsabs.harvard.edu/abs/2023MNRAS.522.1091L}
}

@ARTICLE{bib:Liesenborgs2006,
   author = {{Liesenborgs}, J. and {De Rijcke}, S. and {Dejonghe}, H.},
    title = "{A genetic algorithm for the non-parametric inversion of strong lensing systems}",
  journal = {\mnras},
   eprint = {astro-ph/0601124},
     year = 2006,
    month = apr,
   volume = 367,
    pages = {1209-1216},
      doi = {10.1111/j.1365-2966.2006.10040.x},
   adsurl = {http://adsabs.harvard.edu/abs/2006MNRAS.367.1209L}
}

@MISC{bib:Liesenborgs2010,
   author = {{Liesenborgs}, J. and {de Rijcke}, S. and {Dejonghe}, H.},
    title = "{GRALE: A genetic algorithm for the non-parametric inversion of strong lensing systems}",
howpublished = {Astrophysics Source Code Library},
     year = 2010,
    month = nov,
archivePrefix = "ascl",
   eprint = {1011.021},
   adsurl = {http://adsabs.harvard.edu/abs/2010ascl.soft11021L}
}

@ARTICLE{bib:Liesenborgs2020,
       author = {{Liesenborgs}, Jori and {Williams}, Liliya L.~R. and {Wagner}, Jenny and {De Rijcke}, Sven},
        title = "{Extended lens reconstructions with grale: exploiting time-domain, substructural, and weak lensing information}",
      journal = {\mnras},
         year = 2020,
        month = may,
       volume = {494},
       number = {3},
        pages = {3253-3274},
          doi = {10.1093/mnras/staa842},
archivePrefix = {arXiv},
       eprint = {2003.10377},
 primaryClass = {astro-ph.CO},
       adsurl = {https://ui.adsabs.harvard.edu/abs/2020MNRAS.494.3253L}
}

@ARTICLE{bib:Lin2022,
       author = {{Lin}, Joyce and {Wagner}, Jenny and {Griffiths}, Richard E.},
        title = "{Generalized model-independent characterization of strong gravitational lenses VIII. Automated multiband feature detection to constrain local lens properties}",
      journal = {\mnras},
         year = 2022,
        month = dec,
       volume = {517},
       number = {2},
        pages = {1821-1836},
          doi = {10.1093/mnras/stac2576},
archivePrefix = {arXiv},
       eprint = {2207.01630},
 primaryClass = {astro-ph.CO},
       adsurl = {https://ui.adsabs.harvard.edu/abs/2022MNRAS.517.1821L}
}

@ARTICLE{bib:Lin2023,
       author = {{Lin}, Joyce and {Wagner}, Jenny and {Griffiths}, Richard E.},
        title = "{Much ado about no offset - characterizing the anomalous multiple-image configuration and the model-driven displacement between light and mass in the multiplane strong lens Abell 3827}",
      journal = {\mnras},
         year = 2023,
        month = dec,
       volume = {526},
       number = {2},
        pages = {2776-2794},
          doi = {10.1093/mnras/stad2800},
archivePrefix = {arXiv},
       eprint = {2306.11779},
 primaryClass = {astro-ph.CO},
       adsurl = {https://ui.adsabs.harvard.edu/abs/2023MNRAS.526.2776L}
}

@ARTICLE{bib:Lotz2017,
       author = {{Lotz}, J.~M. and {Koekemoer}, A. and {Coe}, D. and {Grogin}, N. and {Capak}, P. and {Mack}, J. and {Anderson}, J. and {Avila}, R. and {Barker}, E.~A. and {Borncamp}, D. and {Brammer}, G. and {Durbin}, M. and {Gunning}, H. and {Hilbert}, B. and {Jenkner}, H. and {Khandrika}, H. and {Levay}, Z. and {Lucas}, R.~A. and {MacKenty}, J. and {Ogaz}, S. and {Porterfield}, B. and {Reid}, N. and {Robberto}, M. and {Royle}, P. and {Smith}, L.~J. and {Storrie-Lombardi}, L.~J. and {Sunnquist}, B. and {Surace}, J. and {Taylor}, D.~C. and {Williams}, R. and {Bullock}, J. and {Dickinson}, M. and {Finkelstein}, S. and {Natarajan}, P. and {Richard}, J. and {Robertson}, B. and {Tumlinson}, J. and {Zitrin}, A. and {Flanagan}, K. and {Sembach}, K. and {Soifer}, B.~T. and {Mountain}, M.},
        title = "{The Frontier Fields: Survey Design and Initial Results}",
      journal = {\apj},
         year = 2017,
        month = mar,
       volume = {837},
       number = {1},
          eid = {97},
        pages = {97},
          doi = {10.3847/1538-4357/837/1/97},
archivePrefix = {arXiv},
       eprint = {1605.06567},
 primaryClass = {astro-ph.GA},
       adsurl = {https://ui.adsabs.harvard.edu/abs/2017ApJ...837...97L}
}

@ARTICLE{bib:Martin2012,
       author = {{Martin}, Crystal L. and {Shapley}, Alice E. and {Coil}, Alison L. and {Kornei}, Katherine A. and {Bundy}, Kevin and {Weiner}, Benjamin J. and {Noeske}, Kai G. and {Schiminovich}, David},
        title = "{Demographics and Physical Properties of Gas Outflows/Inflows at 0.4 < z < 1.4}",
      journal = {\apj},
         year = 2012,
        month = dec,
       volume = {760},
       number = {2},
          eid = {127},
        pages = {127},
          doi = {10.1088/0004-637X/760/2/127},
archivePrefix = {arXiv},
       eprint = {1206.5552},
 primaryClass = {astro-ph.CO},
       adsurl = {https://ui.adsabs.harvard.edu/abs/2012ApJ...760..127M}
}

@ARTICLE{bib:Massey2015,
       author = {{Massey}, Richard and {Williams}, Liliya and {Smit}, Renske and {Swinbank}, Mark and {Kitching}, Thomas D. and {Harvey}, David and {Jauzac}, Mathilde and {Israel}, Holger and {Clowe}, Douglas and {Edge}, Alastair and {Hilton}, Matt and {Jullo}, Eric and {Leonard}, Adrienne and {Liesenborgs}, Jori and {Merten}, Julian and {Mohammed}, Irshad and {Nagai}, Daisuke and {Richard}, Johan and {Robertson}, Andrew and {Saha}, Prasenjit and {Santana}, Rebecca and {Stott}, John and {Tittley}, Eric},
        title = "{The behaviour of dark matter associated with four bright cluster galaxies in the 10 kpc core of Abell 3827}",
      journal = {\mnras},
         year = 2015,
        month = jun,
       volume = {449},
       number = {4},
        pages = {3393-3406},
          doi = {10.1093/mnras/stv467},
archivePrefix = {arXiv},
       eprint = {1504.03388},
 primaryClass = {astro-ph.CO},
       adsurl = {https://ui.adsabs.harvard.edu/abs/2015MNRAS.449.3393M}
}

@ARTICLE{bib:Massey2018,
       author = {{Massey}, Richard and {Harvey}, David and {Liesenborgs}, Jori and {Richard}, Johan and {Stach}, Stuart and {Swinbank}, Mark and {Taylor}, Peter and {Williams}, Liliya and {Clowe}, Douglas and {Courbin}, Fr{\'e}d{\'e}ric and {Edge}, Alastair and {Israel}, Holger and {Jauzac}, Mathilde and {Joseph}, R{\'e}my and {Jullo}, Eric and {Kitching}, Thomas D. and {Leonard}, Adrienne and {Merten}, Julian and {Nagai}, Daisuke and {Nightingale}, James and {Robertson}, Andrew and {Romualdez}, Luis Javier and {Saha}, Prasenjit and {Smit}, Renske and {Tam}, Sut-Ieng and {Tittley}, Eric},
        title = "{Dark matter dynamics in Abell 3827: new data consistent with standard cold dark matter}",
      journal = {\mnras},
         year = 2018,
        month = jun,
       volume = {477},
       number = {1},
        pages = {669-677},
          doi = {10.1093/mnras/sty630},
archivePrefix = {arXiv},
       eprint = {1708.04245},
 primaryClass = {astro-ph.CO},
       adsurl = {https://ui.adsabs.harvard.edu/abs/2018MNRAS.477..669M}
}

@misc{bib:Newville2025,
  author       = {Newville, Matthew and Otten, Renee and Nelson, Andrew and Stensitzki, Till and Ingargiola, Antonino and Allan, Daniel and Fox, Austin and Carter, Faustin and Rawlik, Michal},
  title        = {LMFIT: Non-Linear Least-Squares Minimization and Curve-Fitting for Python},
  month        = jul,
  year         = 2025,
  publisher    = {Zenodo},
  version      = {1.3.4},
  doi          = {10.5281/zenodo.16175987},
  url          = {https://doi.org/10.5281/zenodo.16175987},
}

@article{bib:pypeit:joss_pub,
    doi = {10.21105/joss.02308},
    url = {https://doi.org/10.21105/joss.02308},
    year = {2020},
    publisher = {The Open Journal},
    volume = {5},
    number = {56},
    pages = {2308},
    author = {J. Xavier Prochaska and Joseph F. Hennawi and Kyle B. Westfall and Ryan J. Cooke and Feige Wang and Tiffany Hsyu and Frederick B. Davies and Emanuele Paolo Farina and Debora Pelliccia},
    title = {PypeIt: The Python Spectroscopic Data Reduction Pipeline},
    journal = {Journal of Open Source Software}
}

@MISC{bib:pypeit:zenodo,
       author = {{Prochaska}, J. Xavier and {Hennawi}, Joseph and {Cooke}, Ryan and
         {Westfall}, Kyle and {Wang}, Feige and {EmAstro} and {Tiffanyhsyu} and
         {Wasserman}, Asher and {Villaume}, Alexa and {Marijana777} and
         {Schindler}, JT and {Young}, David and {Simha}, Sunil and
         {Wilde}, Matt and {Tejos}, Nicolas and {Isbell}, Jacob and
         {Fl{\"o}rs}, Andreas and {Sandford}, Nathan and {Vasovi{\'c}}, Zlatan and
         {Betts}, Edward and {Holden}, Brad},
        title = "{pypeit/PypeIt: Release 1.0.0}",
         year = 2020,
        month = apr,
          eid = {10.5281/zenodo.3743493},
          doi = {10.5281/zenodo.3743493},
      version = {v1.0.0},
    publisher = {Zenodo},
       adsurl = {https://ui.adsabs.harvard.edu/abs/2020zndo...3743493P}
}

@ARTICLE{bib:Richard2014,
       author = {{Richard}, Johan and {Jauzac}, Mathilde and {Limousin}, Marceau and {Jullo}, Eric and {Cl{\'e}ment}, Benjamin and {Ebeling}, Harald and {Kneib}, Jean-Paul and {Atek}, Hakim and {Natarajan}, Priya and {Egami}, Eiichi and {Livermore}, Rachael and {Bower}, Richard},
        title = "{Mass and magnification maps for the Hubble Space Telescope Frontier Fields clusters: implications for high-redshift studies}",
      journal = {\mnras},
         year = 2014,
        month = oct,
       volume = {444},
       number = {1},
        pages = {268-289},
          doi = {10.1093/mnras/stu1395},
archivePrefix = {arXiv},
       eprint = {1405.3303},
 primaryClass = {astro-ph.CO},
       adsurl = {https://ui.adsabs.harvard.edu/abs/2014MNRAS.444..268R}
}

@ARTICLE{bib:Rykoff2016,
       author = {{Rykoff}, E.~S. and {Rozo}, E. and {Hollowood}, D. and {Bermeo-Hernandez}, A. and {Jeltema}, T. and {Mayers}, J. and {Romer}, A.~K. and {Rooney}, P. and {Saro}, A. and {Vergara Cervantes}, C. and {Wechsler}, R.~H. and {Wilcox}, H. and {Abbott}, T.~M.~C. and {Abdalla}, F.~B. and {Allam}, S. and {Annis}, J. and {Benoit-L{\'e}vy}, A. and {Bernstein}, G.~M. and {Bertin}, E. and {Brooks}, D. and {Burke}, D.~L. and {Capozzi}, D. and {Carnero Rosell}, A. and {Carrasco Kind}, M. and {Castander}, F.~J. and {Childress}, M. and {Collins}, C.~A. and {Cunha}, C.~E. and {D'Andrea}, C.~B. and {da Costa}, L.~N. and {Davis}, T.~M. and {Desai}, S. and {Diehl}, H.~T. and {Dietrich}, J.~P. and {Doel}, P. and {Evrard}, A.~E. and {Finley}, D.~A. and {Flaugher}, B. and {Fosalba}, P. and {Frieman}, J. and {Glazebrook}, K. and {Goldstein}, D.~A. and {Gruen}, D. and {Gruendl}, R.~A. and {Gutierrez}, G. and {Hilton}, M. and {Honscheid}, K. and {Hoyle}, B. and {James}, D.~J. and {Kay}, S.~T. and {Kuehn}, K. and {Kuropatkin}, N. and {Lahav}, O. and {Lewis}, G.~F. and {Lidman}, C. and {Lima}, M. and {Maia}, M.~A.~G. and {Mann}, R.~G. and {Marshall}, J.~L. and {Martini}, P. and {Melchior}, P. and {Miller}, C.~J. and {Miquel}, R. and {Mohr}, J.~J. and {Nichol}, R.~C. and {Nord}, B. and {Ogando}, R. and {Plazas}, A.~A. and {Reil}, K. and {Sahl{\'e}n}, M. and {Sanchez}, E. and {Santiago}, B. and {Scarpine}, V. and {Schubnell}, M. and {Sevilla-Noarbe}, I. and {Smith}, R.~C. and {Soares-Santos}, M. and {Sobreira}, F. and {Stott}, J.~P. and {Suchyta}, E. and {Swanson}, M.~E.~C. and {Tarle}, G. and {Thomas}, D. and {Tucker}, D. and {Uddin}, S. and {Viana}, P.~T.~P. and {Vikram}, V. and {Walker}, A.~R. and {Zhang}, Y. and {DES Collaboration}},
        title = "{The RedMaPPer Galaxy Cluster Catalog From DES Science Verification Data}",
      journal = {\apjs},
         year = 2016,
        month = may,
       volume = {224},
       number = {1},
          eid = {1},
        pages = {1},
          doi = {10.3847/0067-0049/224/1/1},
archivePrefix = {arXiv},
       eprint = {1601.00621},
 primaryClass = {astro-ph.CO},
       adsurl = {https://ui.adsabs.harvard.edu/abs/2016ApJS..224....1R}
}

@ARTICLE{bib:Shapley2006,
       author = {{Shapley}, Alice E. and {Steidel}, Charles C. and {Pettini}, Max and {Adelberger}, Kurt L. and {Erb}, Dawn K.},
        title = "{The Direct Detection of Lyman Continuum Emission from Star-forming Galaxies at z\raisebox{-0.5ex}\textasciitilde3}",
      journal = {\apj},
         year = 2006,
        month = nov,
       volume = {651},
       number = {2},
        pages = {688-703},
          doi = {10.1086/507511},
archivePrefix = {arXiv},
       eprint = {astro-ph/0606635},
 primaryClass = {astro-ph},
       adsurl = {https://ui.adsabs.harvard.edu/abs/2006ApJ...651..688S}
}

@ARTICLE{bib:Sharon2012,
       author = {{Sharon}, Keren and {Gladders}, Michael D. and {Rigby}, Jane R. and {Wuyts}, Eva and {Koester}, Benjamin P. and {Bayliss}, Matthew B. and {Barrientos}, L. Felipe},
        title = "{Source-plane Reconstruction of the Bright Lensed Galaxy RCSGA 032727-132609}",
      journal = {\apj},
         year = 2012,
        month = feb,
       volume = {746},
       number = {2},
          eid = {161},
        pages = {161},
          doi = {10.1088/0004-637X/746/2/161},
archivePrefix = {arXiv},
       eprint = {1202.0539},
 primaryClass = {astro-ph.CO},
       adsurl = {https://ui.adsabs.harvard.edu/abs/2012ApJ...746..161S}
}

@ARTICLE{bib:Sharon2022,
       author = {{Sharon}, Keren and {Cerny}, Catherine and {Rigby}, Jane R. and {Florian}, Michael K. and {Bayliss}, Matthew B. and {Dahle}, Hakon and {Gladders}, Michael D. and {Mahler}, Guillaume},
        title = "{HST-Based Lens Model of SDSS J1226+2152, in Preparation for JWST-ERS TEMPLATES}",
      journal = {arXiv e-prints},
         year = 2022,
        month = jul,
          eid = {arXiv:2207.05709},
        pages = {arXiv:2207.05709},
          doi = {10.48550/arXiv.2207.05709},
archivePrefix = {arXiv},
       eprint = {2207.05709},
 primaryClass = {astro-ph.GA},
       adsurl = {https://ui.adsabs.harvard.edu/abs/2022arXiv220705709S}
}

@ARTICLE{bib:Sheu2024,
       author = {{Sheu}, William and {Cikota}, Aleksandar and {Huang}, Xiaosheng and {Glazebrook}, Karl and {Storfer}, Christopher and {Agarwal}, Shrihan and {Schlegel}, David J. and {Suzuki}, Nao and {Barone}, Tania M. and {Bian}, Fuyan and {Jeltema}, Tesla and {Jones}, Tucker and {Kacprzak}, Glenn G. and {O'Donnell}, Jackson H. and {G.~C.}, Keerthi Vasan},
        title = "{The Carousel Lens: A Well-modeled Strong Lens with Multiple Sources Spectroscopically Confirmed by VLT/MUSE}",
      journal = {\apj},
         year = 2024,
        month = sep,
       volume = {973},
       number = {1},
          eid = {3},
        pages = {3},
          doi = {10.3847/1538-4357/ad65d3},
archivePrefix = {arXiv},
       eprint = {2408.10320},
 primaryClass = {astro-ph.GA},
       adsurl = {https://ui.adsabs.harvard.edu/abs/2024ApJ...973....3S}
}

@ARTICLE{bib:Steidel2003,
       author = {{Steidel}, Charles C. and {Adelberger}, Kurt L. and {Shapley}, Alice E. and {Pettini}, Max and {Dickinson}, Mark and {Giavalisco}, Mauro},
        title = "{Lyman Break Galaxies at Redshift z \raisebox{-0.5ex}\textasciitilde 3: Survey Description and Full Data Set}",
      journal = {\apj},
         year = 2003,
        month = aug,
       volume = {592},
       number = {2},
        pages = {728-754},
          doi = {10.1086/375772},
archivePrefix = {arXiv},
       eprint = {astro-ph/0305378},
 primaryClass = {astro-ph},
       adsurl = {https://ui.adsabs.harvard.edu/abs/2003ApJ...592..728S}
}

@ARTICLE{bib:Uzsoy2025,
       author = {{Uzsoy}, Ana Sof{\'\i}a M. and {Dey}, Arjun and {Raichoor}, Anand and {Finkbeiner}, Douglas P. and {Ramakrishnan}, Vandana and {Lee}, Kyoung-Soo and {Gawiser}, Eric and {Aguilar}, Jessica Nicole and {Ahlen}, Steven and {Anand}, Abhijeet and {Bianchi}, Davide and {Brooks}, David and {Claybaugh}, Todd and {de la Macorra}, Axel and {Doel}, Peter and {Ferraro}, Simone and {Firestone}, Nicole M. and {Font-Ribera}, Andreu and {Forero-Romero}, Jaime E. and {Gazta{\~n}aga}, Enrique and {Guaita}, Lucia and {Gutierrez}, Gaston and {Herrera-Alcantar}, Hiram K. and {Hwang}, Ho Seong and {Ishak}, Mustapha and {Joyce}, Dick and {Kirkby}, David and {Kisner}, Theodore and {Kremin}, Anthony and {Lahav}, Ofer and {Lamman}, Claire and {Landriau}, Martin and {Le Guillou}, Laurent and {Manera}, Marc and {Miquel}, Ramon and {Moustakas}, John and {Mu{\~n}oz-Guti{\'e}rrez}, Andrea and {Nadathur}, Seshadri and {Palanque-Delabrouille}, Nathalie and {Percival}, Will and {Poppett}, Claire and {Prada}, Francisco and {P{\'e}rez-R{\`a}fols}, Ignasi and {Rossi}, Graziano and {Sanchez}, Eusebio and {Schlegel}, David and {Schubnell}, Michael and {Seo}, Hee-Jong and {Silber}, Joseph Harry and {Song}, Hyunmi and {Sprayberry}, David and {Tarl{\'e}}, Gregory and {Weaver}, Benjamin Alan and {Zou}, Hu},
        title = "{Effect of local environment on Ly$α$ line profile in DESI/ODIN LAEs}",
      journal = {arXiv e-prints},
         year = 2025,
        month = nov,
          eid = {arXiv:2511.17498},
        pages = {arXiv:2511.17498},
archivePrefix = {arXiv},
       eprint = {2511.17498},
 primaryClass = {astro-ph.GA},
       adsurl = {https://ui.adsabs.harvard.edu/abs/2025arXiv251117498U}
}

@ARTICLE{bib:Vanzella2021,
       author = {{Vanzella}, E. and {Caminha}, G.~B. and {Rosati}, P. and {Mercurio}, A. and {Castellano}, M. and {Meneghetti}, M. and {Grillo}, C. and {Sani}, E. and {Bergamini}, P. and {Calura}, F. and {Caputi}, K. and {Cristiani}, S. and {Cupani}, G. and {Fontana}, A. and {Gilli}, R. and {Grazian}, A. and {Gronke}, M. and {Mignoli}, M. and {Nonino}, M. and {Pentericci}, L. and {Tozzi}, P. and {Treu}, T. and {Balestra}, I. and {Dijkstra}, M.},
        title = "{The MUSE Deep Lensed Field on the Hubble Frontier Field MACS J0416. Star-forming complexes at cosmological distances}",
      journal = {\aap},
         year = 2021,
        month = feb,
       volume = {646},
          eid = {A57},
        pages = {A57},
          doi = {10.1051/0004-6361/202039466},
archivePrefix = {arXiv},
       eprint = {2009.08458},
 primaryClass = {astro-ph.GA},
       adsurl = {https://ui.adsabs.harvard.edu/abs/2021A&A...646A..57V}
}

@ARTICLE{bib:Verhamme2017,
       author = {{Verhamme}, A. and {Orlitov{\'a}}, I. and {Schaerer}, D. and {Izotov}, Y. and {Worseck}, G. and {Thuan}, T.~X. and {Guseva}, N.},
        title = "{Lyman-{\ensuremath{\alpha}} spectral properties of five newly discovered Lyman continuum emitters}",
      journal = {\aap},
         year = 2017,
        month = jan,
       volume = {597},
          eid = {A13},
        pages = {A13},
          doi = {10.1051/0004-6361/201629264},
archivePrefix = {arXiv},
       eprint = {1609.03477},
 primaryClass = {astro-ph.GA},
       adsurl = {https://ui.adsabs.harvard.edu/abs/2017A&A...597A..13V}
}

@ARTICLE{bib:Wagner1,
   author = {{Wagner}, J.},
    title = "{Generalised model-independent characterisation of strong gravitational lenses. I. Theoretical foundations}",
  journal = {\aap},
archivePrefix = "arXiv",
     year = 2017,
   volume = 601,
      eid = {A131},
    pages = {A131},
      doi = {10.1051/0004-6361/201630200},
   adsurl = {http://cdsads.u-strasbg.fr/abs/2017A%26A...601A.131W}
}

@ARTICLE{bib:Wagner2,
   author = {{Wagner}, J. and {Tessore}, N.},
    title = "{Generalised model-independent characterisation of strong gravitational lenses. II. Transformation matrix between multiple images}",
  journal = {\aap},
archivePrefix = "arXiv",
     year = 2018,
   volume = 613,
      eid = {A6},
    pages = {A6},
      doi = {10.1051/0004-6361/201730947},
   adsurl = {http://cdsads.u-strasbg.fr/abs/2018A%26A...613A...6W}
}

@ARTICLE{bib:Wagner3,
   author = {{Wagner}, J.},
    title = "{Generalised model-independent characterisation of strong gravitational lenses. III. Perturbed axisymmetric lenses}",
  journal = {\aap},
archivePrefix = "arXiv",
     year = 2018,
   volume = 615,
      eid = {A102},
    pages = {A102},
      doi = {10.1051/0004-6361/201731207},
   adsurl = {http://cdsads.u-strasbg.fr/abs/2018A%26A...615A.102W}
}

@ARTICLE{bib:Wagner4,
   author = {{Wagner}, J.},
    title = "{Generalised model-independent characterisation of strong gravitational lenses. IV. Formalism-intrinsic degeneracies}",
  journal = {\aap},
archivePrefix = "arXiv",
     year = 2018,
   volume = 620,
      eid = {A86},
    pages = {A86},
      doi = {10.1051/0004-6361/201834218},
   adsurl = {http://cdsads.u-strasbg.fr/abs/2018A%26A...620A..86W}
}

@ARTICLE{bib:Wagner5,
       author = {{Wagner}, Jenny and {Meyer}, Sven},
        title = "{Generalized model-independent characterization of strong gravitational lenses V: reconstructing the lensing distance ratio by supernovae for a general Friedmann universe}",
      journal = {\mnras},
         year = 2019,
        month = dec,
       volume = {490},
       number = {2},
        pages = {1913-1927},
          doi = {10.1093/mnras/stz2717},
archivePrefix = {arXiv},
       eprint = {1812.04002},
 primaryClass = {astro-ph.CO},
       adsurl = {https://ui.adsabs.harvard.edu/abs/2019MNRAS.490.1913W}
}

@ARTICLE{bib:Wagner6,
       author = {{Wagner}, Jenny},
        title = "{Generalised model-independent characterisation of strong gravitational lenses - VI. The origin of the formalism intrinsic degeneracies and their influence on H$_{0}$}",
      journal = {\mnras},
         year = "2019",
       volume = {487},
       number = {4},
        pages = {4492-4503},
          doi = {10.1093/mnras/stz1587},
archivePrefix = {arXiv},
       eprint = {1904.07239},
 primaryClass = {astro-ph.CO},
       adsurl = {https://ui.adsabs.harvard.edu/abs/2019MNRAS.487.4492W}
}

@ARTICLE{bib:Wagner7,
       author = {{Wagner}, Jenny},
        title = "{Generalised model-independent characterisation of strong gravitational lenses. VII. Impact of source properties and higher-order lens properties on the local lens reconstruction}",
      journal = {\aap},
         year = 2022,
        month = jul,
       volume = {663},
          eid = {A157},
        pages = {A157},
          doi = {10.1051/0004-6361/202243562},
archivePrefix = {arXiv},
       eprint = {2203.06190},
 primaryClass = {astro-ph.CO},
       adsurl = {https://ui.adsabs.harvard.edu/abs/2022A&A...663A.157W}
}

@ARTICLE{bib:Wagner_cluster0024,
       author = {{Wagner}, Jenny and {Liesenborgs}, Jori and {Tessore}, Nicolas},
        title = "{Model-independent and model-based local lensing properties of CL0024+1654 from multiply imaged galaxies}",
      journal = {Astronomy \& Astrophysics},
         year = 2018,
        month = apr,
       volume = {612},
          eid = {A17},
        pages = {A17},
          doi = {10.1051/0004-6361/201731932},
archivePrefix = {arXiv},
       eprint = {1709.03531},
 primaryClass = {astro-ph.CO},
       adsurl = {https://ui.adsabs.harvard.edu/abs/2018A&A...612A..17W}
}

@ARTICLE{bib:Wagner_quasar,
       author = {{Wagner}, Jenny and {Williams}, Liliya L.~R.},
        title = "{Model-independent and model-based local lensing properties of B0128+437 from resolved quasar images}",
      journal = {\aap},
         year = 2020,
        month = mar,
       volume = {635},
          eid = {A86},
        pages = {A86},
          doi = {10.1051/0004-6361/201936628},
archivePrefix = {arXiv},
       eprint = {1909.01349},
 primaryClass = {astro-ph.GA},
       adsurl = {https://ui.adsabs.harvard.edu/abs/2020A&A...635A..86W}
}

@ARTICLE{bib:Williams2011,
       author = {{Williams}, Liliya L.~R. and {Saha}, Prasenjit},
        title = "{Light/mass offsets in the lensing cluster Abell 3827: evidence for collisional dark matter?}",
      journal = {\mnras},
         year = 2011,
        month = jul,
       volume = {415},
       number = {1},
        pages = {448-460},
          doi = {10.1111/j.1365-2966.2011.18716.x},
archivePrefix = {arXiv},
       eprint = {1102.3943},
 primaryClass = {astro-ph.CO},
       adsurl = {https://ui.adsabs.harvard.edu/abs/2011MNRAS.415..448W}
}

\begin{appendix}
\section{From raw data to the data cube}
\label{sec:data_reduction}

\subsection{Data Acquisition}
\label{sec:kcwi_data_acquisition}

The data were taken with KCWI at the right Nasmyth focus of the Keck II telescope on September 16, 2020 under the programme ID H315 with Richard E.~Griffiths as Principal Investigator.  
The configuration was set to a 'Large' slicer (1.35~arcsec slice width) and a field of view of $33$~arcsec~$\times~20.4$~arcsec with a 'BL' grating centred at 4500~\AA, R(central)=900, a bandpass/dispersion of 0.563\AA/pixel and no red (blocking) filter. 
The effective spectral coverage was $\lambda \in \left[ 3500, 5600 \right]$~\AA.
Orienting the long dimension of the IFU in North-South direction, all three multiple images were covered by a single field centred around RA=337.540083~deg and Dec=-8.1583~deg. 
The atmospheric seeing was typically 0.3~arcsec and thus better than the spatial sampling of the final data cube, 1.35~arcsec. 

The data are publicly available on the W.~M.~Keck Observatory Archive (KOA) server\footnote{\url{https://koa.ipac.caltech.edu/cgi-bin/KOA/nph-KOAlogin}}. 
The observations encompass 16 science exposures: two observations of a standard star, one with 60s and 120s exposure time and 14 observations of the target triple-image configuration, split into four with 900s and 10 with 1200s exposure times. 
In addition, 48 calibration files were obtained: 
9 focus bias frames, which were not used by our data reduction, 
7 bias frames, which were used by our data reduction,
6 frames in the dark-dome mode to account for spatial variations in slit illumination as a result of vignetting or instrument optics and to perform a slit edge detection,
12 frames in the pixel-flat mode to account for pixel-to-pixel differences in the quantum efficiency of the detector, 
2 frames for astrometric alignment, 
2 frames to correct for a tilt in the observations employing a Thorium-Argon lamp for this calibration observation, 
2 arc frames generated with an Iron-Argon lamp for the wavelength calibration, 
8 twilight flat frames with two different pointings on the sky to perform a relative scale correction when co-adding all science frames.
The latter ensures that the relative spectral sensitivity of the joint data cube is constant across the field of view.

\subsection{Data Calibration and Reduction}
\label{sec:kcwi_data_calibration}

To reduce the science frames, we used \texttt{PypeIt}, a Python package for semi-automated reduction of astronomical spectroscopic data, developed and made publicly available in \cite{bib:pypeit:joss_pub} and \cite{bib:pypeit:zenodo}. 
It already contains many KCWI specific processing steps\footnote{\url{https://pypeit.readthedocs.io/en/latest/spectrographs/keck_kcwi.html}}, so that the data were reduced without much manual fine-tuning. 
 
Setting up the configuration file for the data reduction pipeline, we followed all standard steps recommended for KCWI, especially, we used the FeAr lamp for wavelength calibration and the ThAr lamp for tilting and we kept the pixel-flat correction and the astrometric alignment active.
We also kept the default removal of the sinusoidal CCD pattern switched on, as well as the subtraction of scattered light. 
As the spectral and spatial flexure corrections required a good sky detection, we did not perform these steps because we deferred the sky subtraction to the final co-addition of all calibrated and reduced science frames at the end of the process. 

Additional modifications to the standard \texttt{PypeIt} configuration file are:
we followed the manual and did not clip the arc frames and tilt-correction frames, yet, subtracted the continuum from both to obtain a reduced variance in the wavelength calibrations compared to the standard procedure of switching the continuum subtraction off.  
With the aim of joining all science exposures into a single data cube in the next step and only calibrate their spectral sensitivity together, we switched off the spectral illumination correction when reducing the individual science frames. 

\begin{figure}
\centering
\includegraphics[width=\linewidth]{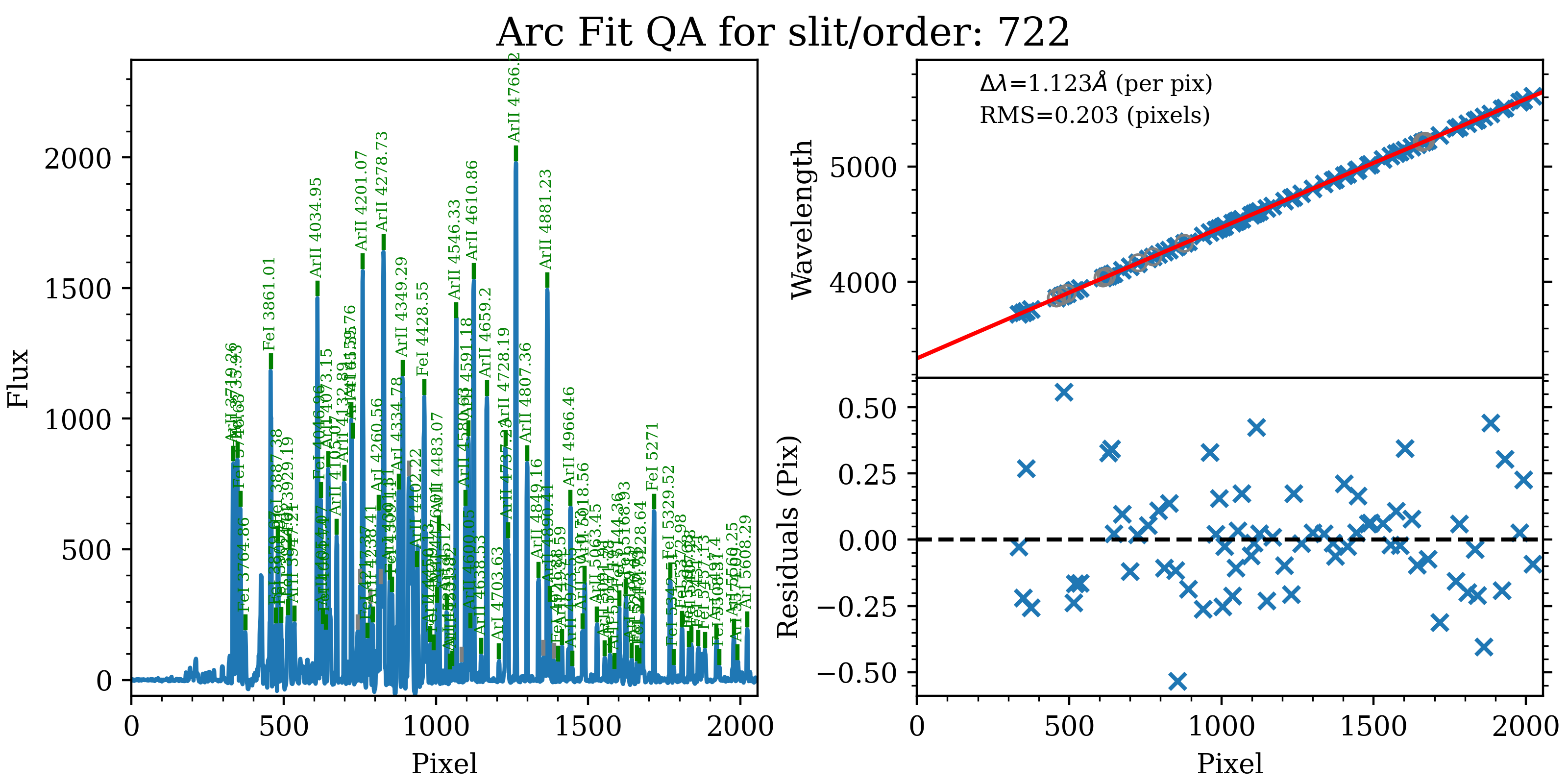}
\caption{Wavelength calibration performed by \texttt{PypeIt}: Typical quality assessment plot for one slit in which the identified FeAr emission lines are marked on top of the measured spectrum (left plot). Then, these lines and their known wavelength differences are used to convert the pixel-based spectrum into a wavelength-based one (right plot). (The fitted function is a Legendre polynomial of order 1 by default, applied to each slit. The joint wavelength function employing all slits is of order 4.)}
\label{fig:wavelength_calibration}
\end{figure}

To check the quality of the data reduction, \texttt{PypeIt} generates several quality assessment files and we scrutinised the outputs according to the \texttt{PypeIt} manual, for instance, checking the correct identification of the slit edges, the visual appearance of the bias frame, the arc and tilt frames, and the flat frames. 

For the wavelength calibration, we inspected the plots like Fig.~\ref{fig:wavelength_calibration} that show the spectral lines emitted by the FeAr lamp identified by the pipeline (left). 
The upper right plot shows the fit which converts pixel into wavelength intervals, including the quality-of-fit values, the lower right plot shows the residuals. 
While a good fit is expected to have an RMS below 0.1 according to the manual, all RMS-values in our data cube were between 0.203 and 0.277. 
Fine-tuning the parameters to improve the fits, we let \texttt{PypeIt} identify the emission lines with their individual widths (instead of a default fixed width) and adjusted the Legendre polynomial fit for each slit to be of order 3 instead of 1. 
We also tried to exclude poorly identified lines by decreasing the tolerance when matching the measured lines with the known wavelength of each line from 2.0 pixels to 1.5 pixels, without significant improvements in the fits. 
Another test included re-running the pipeline without subtracting the continuum in the arc and tilt frames, which yielded larger RMS-values for the wavelength calibrations of the individual slits.

Given that plenty of lines had been identified without a visible bias in the residuals, we continued to the next step and accounted for the increased scatter by an increased uncertainty in the wavelength determination of emission and absorption lines of our target galaxy. 
As will become evident in Section~\ref{sec:kcwi_multiple-image_identification}, the uncertainties in the location of the multiple-images within the reduced data cube are larger than the one caused by the sub-optimal wavelength calibration.

As a final result, we obtained 14 reduced science frames covering the targeted triple-image configuration that contained the pre-processed data. 
In the next step, these data were joined into a data cube covering the sky area of the triple-image configuration and, for each pixel, having a spectrum with $\lambda \in \left[ 3500, 5600\right]$~\AA.

\subsection{Data Cube Generation}
\label{sec:kcwi_data_cube}

Before co-adding the reduced science files to the final 3d data cube, we performed a relative spectral illumination calibration by first reducing the six sky frames with the same pointing onto the sky like science frames.
The only difference was that the sky subtraction part was performed jointly, meaning under the assumption that the sky was the same for all six frames and that the sky was covering the entire frame. 
The frame with the longest exposure time produced after this reduction step was then inserted into the co-adding configuration file to perform a relative spectral sensitivity scaling. 

The co-addition itself was performed such that the obtained spectrum for each sky-pixel is plotted with respect to the observed wavelength, i.e.~without making any assumption about the redshifts of the objects contained in the pixels. 
We restricted the wavelengths to $\lambda \in \left[ 3500, 5500\right]$~\AA \ for stability reasons. 
The wavelength calibration yielded 1.12~\AA \ per spectral pixel, which will be the limiting precision to determine the maxima and minima of the spectral lines.

As science frames with different exposure times are not recommended to be co-added, we restricted ourselves to the 10 frames with 1200~s each.
Out of these, three could not be included due to processing errors, lacking any unmasked pixels inside. 
Thus, the final data cube is a co-addition of 2.33~hours of total exposure time. 
Alongside the data cube, the co-addition also produced a white-light image of the data cube, see Fig.~\ref{fig:cube_fit1} (left). 
The latter was used to identify the pixels containing the multiple images and to extract their spectra, as described in the next section.

During the data reduction, the pipeline was not able to extract any one-dimensional spectrum of a reference star, so that no fluxing could be performed. 
As a consequence, all spectra carry units of counts instead of fluxes. 
Yet, since we focus on the wavelengths of the emission and absorption features and the relative amplitudes of emission and absorption lines for each spectrum individually, a flux calibration is not required to determine the redshift.

\onecolumn
\section{Additional plots}
\label{sec:additional_plots}

Fig.~\ref{fig:mgii} shows a detail of the spectra of all three multiple images around the MgII doublet (left) and the last absorption feature (right) including (dark coloured lines) and excluding the star-forming region in the outskirts of the galaxy (bright coloured lines).

\begin{figure*}[h!]
\centering
\includegraphics[width=0.46\linewidth]{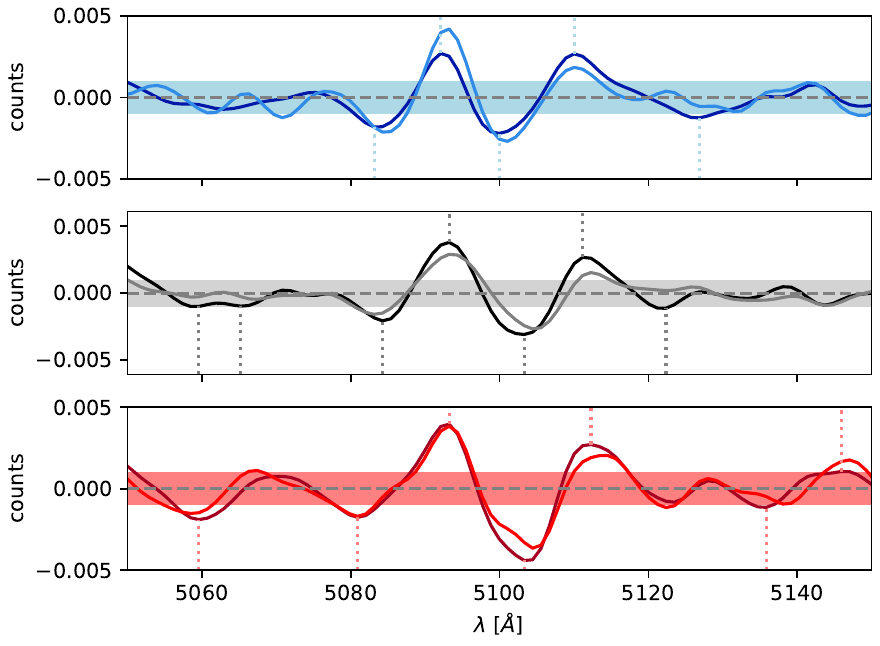} \hfill
\includegraphics[width=0.53\linewidth]{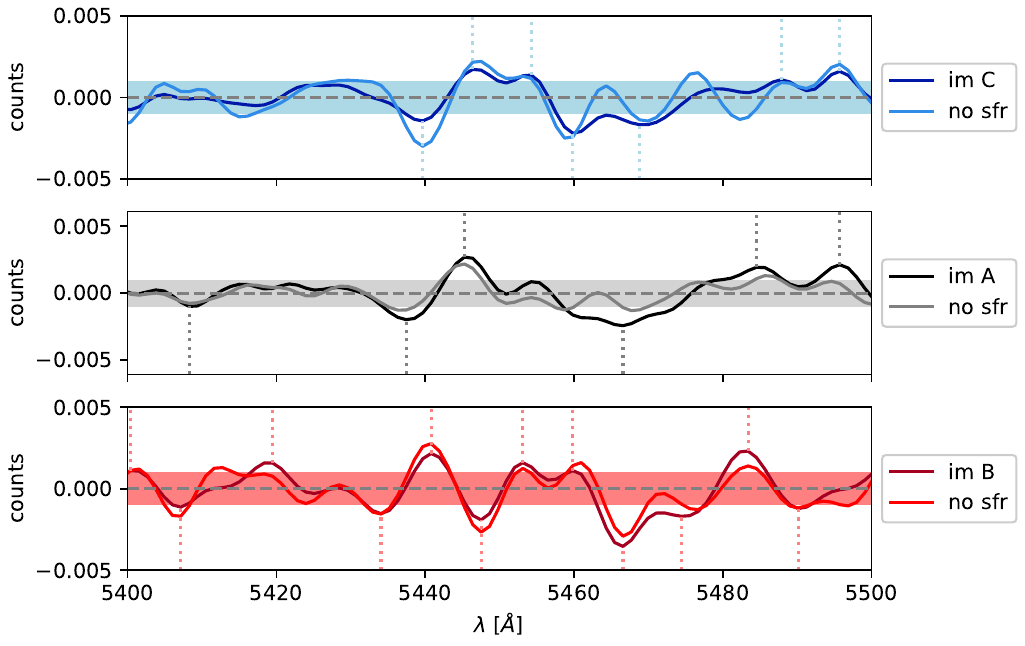}
\caption{Detail of Fig.~\ref{fig:mis_spectra_nosfr} to zoom into the MgII doublet (left) and into the last absorption feature (right). Dotted lines to the top mark emission lines above 3 $\sigma$ signal-to-noise as detected in each spectrum individually, analogously for the absorption lines that are marked by dotted lines to the bottom of the plot.}
\label{fig:mgii}
\end{figure*}

Fig.~\ref{fig:4500-5100} shows the spectra of all three multiple images in the wavelength range $\lambda = \left[4500,5100 \right]$~\AA \ as a detail plot of Fig.~\ref{fig:mis_spectra_nosfr} to study the iron lines in greater detail. 

\begin{figure*}[h!]
\centering
\includegraphics[width=\linewidth]{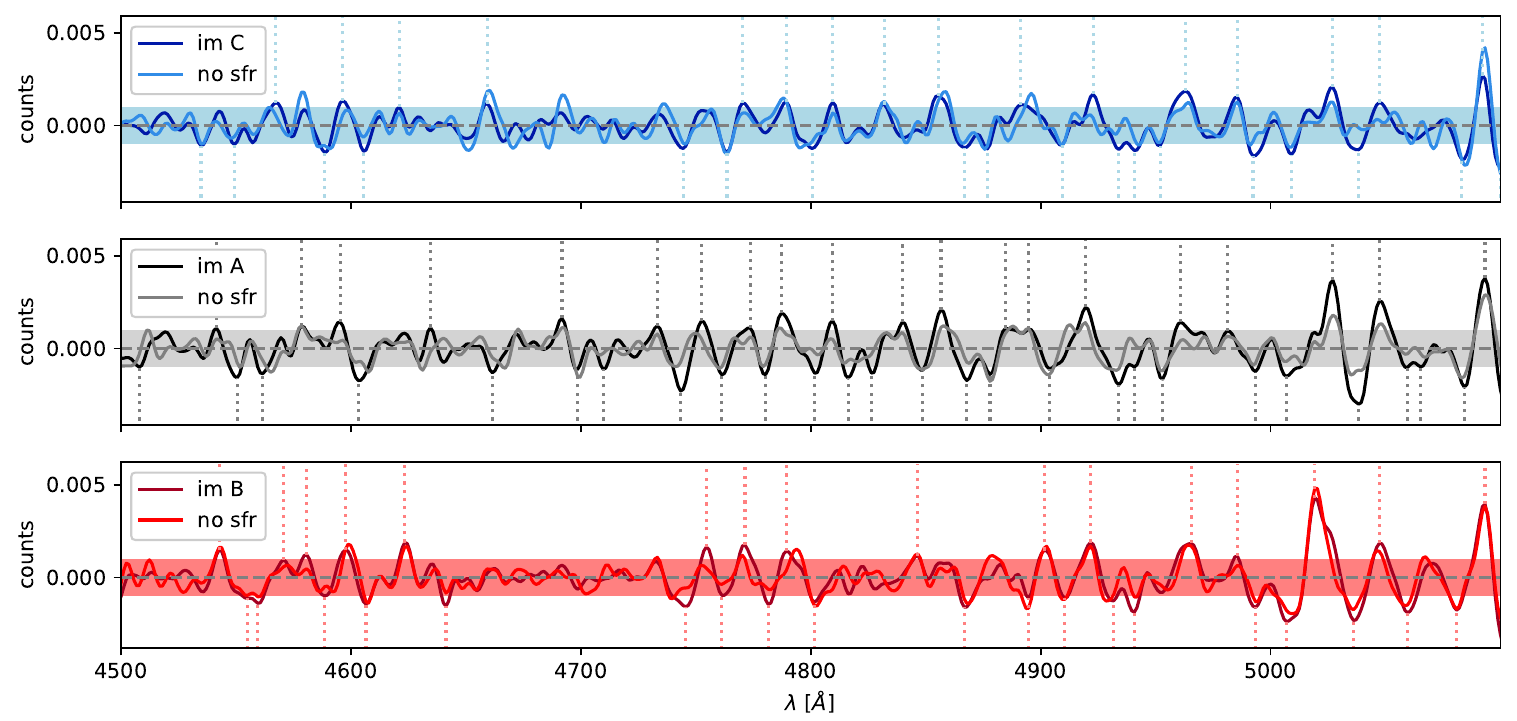}
\caption{Detail of Fig.~\ref{fig:mis_spectra_nosfr} to zoom into the iron line patterns. Dotted lines to the top mark emission lines above 3 $\sigma$ signal-to-noise as detected in each spectrum individually, analogously for the absorption lines that are marked by dotted lines to the bottom of the plot.}
\label{fig:4500-5100}
\end{figure*}

\end{appendix}

\end{document}